\def\be{\begin{equation}}
\def\ee{\end{equation}}
\def\ba{\begin{eqnarray}}
\def\ea{\end{eqnarray}}
\def\nn{\nonumber}

\def\threej#1#2#3#4#5#6{\left( \begin{array}{ccc} #1 & #2 & #3 \\ #4 & #5 & #6 \end{array} \right) }
\def\eqdef{\stackrel{\rm def}{=}}
\def\bigoh{{\mathcal O}}
\def\Cov{\mbox{Cov}}
\def\Var{\mbox{Var}}

\def\Npix{N_{\rm pix}}
\def\Nchan{N_{\rm chan}}
\def\Nside{N_{\rm side}}
\def\Ntmpl{N_{\rm tmpl}}
\def\ellmin{\ell_{\rm min}}
\def\ellmax{\ell_{\rm max}}
\def\ellsplit{\ell_{\rm split}}
\def\fsky{f_{\rm sky}}
\def\simle{\lesssim}
\def\simge{\gtrsim}
\def\tT{\widetilde T}
\def\bn{{\bf \widehat n}}

\def\hC{\widehat C}
\def\DC{\Delta \widehat C}
\def\bd{{\bf d}}
\def\lest{\widehat{\mathcal C}}
\def\ta{\widetilde a}
\def\tg{\widetilde g}
\def\tphi{\widetilde \phi}
\def\tpsi{\widetilde \psi}
\def\ts{\widetilde s}

\def\NF{{\rm 1.4 GHz}}
\def\systematics{\S\ref{sec:nvss_systematics}-\S\ref{sec:sz}}
\def\smax{{s_{\rm smax}}}
\def\e{\hat{\bf e}}

\def\arcdeg{\ensuremath{^{\circ}}}
\def\gau{{\mathcal G}}
\def\hphi{{\widehat \phi}}

\documentclass[twocolumn,amsmath,amssymb,prd]{revtex4}
\usepackage{bm}
\usepackage{epsf}
\usepackage{graphicx}

\begin{document}

\title{Detection of Gravitational Lensing in the Cosmic Microwave Background}

\author{Kendrick M. Smith}
\affiliation{Kavli Institute for Cosmological Physics, University of Chicago, 60637 USA}
\author{Oliver Zahn}
\affiliation{Harvard-Smithsonian Center for Astrophysics, 60 Garden Street, Cambridge, MA 02138 USA}
\author{Olivier Dor\'e}
\affiliation{Canadian Institute for Theoretical Astrophysics, 60 St. George St, University of Toronto, Toronto ON Canada M5S 3H8}

\baselineskip 11pt
\begin{abstract}
Gravitational lensing of the cosmic microwave background (CMB), a long-standing prediction of
the standard cosmolgical model, is ultimately expected to be an important source of cosmological information,
but first detection has not been achieved to date.
We report a 3.4$\sigma$ detection, by applying quadratic estimator techniques to all sky maps from 
the Wilkinson Microwave Anisotropy Probe (WMAP) satellite, and correlating the result with radio galaxy counts 
from the NRAO VLA Sky Survey (NVSS).
We present our methodology including a detailed discussion of potential contaminants.
Our error estimates include systematic uncertainties
from density gradients in NVSS, beam effects in WMAP, Galactic
microwave foregrounds, resolved and unresolved CMB point sources, and
the thermal Sunyaev-Zeldovich effect.
\end{abstract} 

\maketitle

\section{Introduction}

Within just two decades, cosmology has progressed from a rather
speculative science to one of the most successful fields of physics,
driven by an exemplary interplay between experiment and theory. Much
of this progress has been owing to the well understood physics
underlying the Comic Microwave Background (CMB) anisotropy, seeded by
oscillations in the baryon-photon plasma of the early universe.

Measurements of these fluctuations by a number of experiments have
given rise to a basic cosmological paradigm, with the tightest current
constraints on the cosmological parameter budget coming from
combinations of data from the Wilkinson Microwave Anisotropy Probe
(WMAP) satellite \cite{Bennett:2003bz,Spergel:2006hy} in conjunction with small scale
CMB experiments (e.g. \cite{Jones:2005yb,Readhead:2004gy,Kuo:2006ya}), and
other rich probes of cosmological clustering and dynamics such as
supernovae, galaxy surveys, the Lyman-alpha forest, weak lensing, and others (e.g. \cite{Riess:1998cb,Perlmutter:1998np,York:2000gk,Colless:2001gk,Seljak:2006bg,Hoekstra:2002xs,Astier:2005qq,Refregier:2003ct,Tegmark:2006az}).

The CMB promises to remain a gold mine for precision cosmology, and
two new frontiers lie ahead. First, a polarized component has
recently been detected by a number of groups
\cite{Kovac:2002fg,Piacentini:2005yq,Readhead:2004xg,Barkats:2004he,Montroy:2005yx,Johnson:2006jk}, 
offering e.g. the prospects of detecting primordial gravitational waves and
constraining recombination physics. 

Second, large scale structure between the last scattering surface
and us alters the primary CMB anisotropy, through gravitational
lensing (for a recent review of the theory see \cite{Lewis:2006fu}),
through scattering off hot electrons in large scale structure (the
Sunyaev-Zel'dovich effects) \cite{sz70,sz80}, and through redshifting
during the traverse of time-dependent potential fluctuations 
(the ISW effect) \cite{Sachs:1967er}. A number of
specialized instruments will soon begin to study details of these
secondary anisotropies \cite{Kosowsky:2004sw,Ruhl:2004kv}. 

As important as constraining cosmological and astrophysical
parameters, detecting any of these effects is a crucial milestone for
cosmological physics. The Sunyaev-Zel'dovich effect has been found
by targeting clusters detected in X-ray \cite{Birkinshaw:SZdet,Holzapfel:1997kx,Dawson:2002dg,LaRoque:2002si},
also at high significance level using WMAP \cite{Afshordi:2006pc}, and it has been observed in cross-correlation of galaxy surveys with WMAP \cite{Hernandez-Monteagudo:2004sm,Afshordi:2003xu,Fosalba:2003ge}.
The ISW effect has been detected in cross-correlation of WMAP with galaxy surveys and with the hard X-ray background
\cite{Boughn:2003yz,Boughn:2004zm,Nolta:2003uy,Maddox:1990hb,Fosalba:2003ge,Scranton:2003in,Fosalba:2003iy,Padmanabhan:2004fy,Cabre:2006qm,McEwen:2006md}.     

A detection of gravitational lensing in the CMB has so far been outstanding. 
The main difficulty at millimeter wavelengths is the high angular resolution needed, 
as typical deflection angles over a cosmological volume are only a few arcminutes.
Non-Gaussianity imprinted by lensing into the primordial CMB may allow 
statistical detection with surveys at lower angular resolution, but the signal-to-noise is currently 
too low for internal detections. Cross correlation with other tracers of large scale
structure offers a way to limit systematics and increase the signal to noise. 

A first attempt \cite{Hirata:2004rp} was made by cross correlating
the WMAP first year release \cite{Bennett:2003bz} with data from the
Sloan Digital Sky Survey (SDSS) \cite{York:2000gk}. These authors used
a sample of 503,944 SDSS Luminous Red Galaxies (LRG's) overlapping
with $\simeq 10\%$ of the sky observed by WMAP. They were
not able to find evidence for gravitational lensing within statistical
error bounds.  
While SDSS LRG's have
a well understood redshift distribution, their number density drops
rapidly beyond $z=0.5$, and has only marginal overlap with the higher
redshift range that is geometrically optimal for CMB
lensing. Photometric quasars found in SDSS may offer an additional
handle. 

Here we go a different route, using the 1.9 million radio sources
found in the NRAO VLA Sky Survey (NVSS)  \cite{Condon:1998iy}. 
The large sky coverage and estimated depth of NVSS make it an excellent candidate for a 
search for CMB lensing in cross correlation with WMAP. 
The survey covers 77\% of the sky, 58\% of which is found to overlap with WMAP, 
once masks to limit systematics have been applied.

The structure of this paper is as follows. 

First, we describe the datasets (\S\ref{sec:data}), theory (\S\ref{sec:lenstheory}), 
and pipeline (\S\ref{sec:pipeline}) that
will be used for detecting CMB lensing by reconstructing the lensing potential from
WMAP, and cross-correlating the result to NVSS.
The detection is shown, with statistical errors only, in Fig.~\ref{fig:statdetection}.
The rest of the paper is devoted to null tests and assigning systematic errors:
NVSS systematics (\S\ref{sec:nvss_systematics}), 
WMAP beam effects and Galactic foregrounds (\S\ref{sec:wmap_systematics}),
resolved and unresolved point sources (\S\ref{sec:pointsources}),
and Sunyaev-Zeldovich fluctuations (\S\ref{sec:sz}).
We quote our final result including systematic errors (Fig.~\ref{fig:final}) in 
\S\ref{sec:discussion}, where we also mention future directions. 

In our calculations we will assume throughout the cosmological model
favored by a combination of WMAP, smaller scale CMB experiments, and
other data (the WMAP+ALL analysis, \cite{Spergel:2006hy}): a local
expansion rate $H_0=70.4$ km/s/Mpc, primordial power spectrum
slope $n_s=0.947$, matter and dark energy fractions of
$\Omega_0=0.267$ and $\Omega_\Lambda=0.733$ respectively, and
amplitude $\sigma_8=0.773$.

\section{Datasets}
\label{sec:data}

\subsection{WMAP}
\label{ssec:wmap}

\begin{figure}
\centerline{\epsfxsize=3.2truein\epsffile[40 460 550 700]{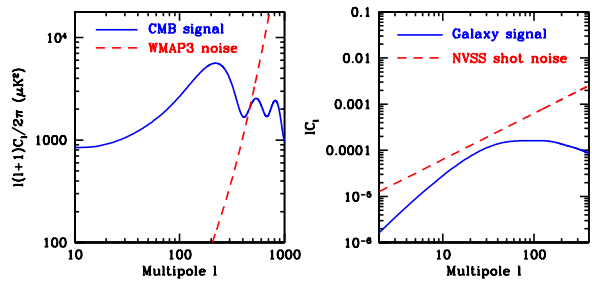}}
\caption{{\bf Left panel:} CMB signal power spectrum, and three-year WMAP noise power spectrum.
{\bf Right panel:} Fiducial NVSS signal power spectrum, and NVSS shot noise 
($\overline G = 159000$ gal/steradian).}
\label{fig:dataset_noise}
\end{figure}

With the goal of producing full sky maps of the CMB with unprecedented
accuracy, the WMAP satellite was launched in June 2001. Since then, it has been mapping the
sky using 10 Differential Assemblies covering 5 frequency bands centered at  23 (K), 33 (Ka), 41 (Q), 61
(V) and	94 GHz (W). In our analysis we use the 2 Q-band, 2 V-band, and 4 W-band temperature
maps produced using 3 years of observations \cite{Hinshaw:2006ia}
and made publicly available \footnote{{\texttt http://lambda.gsfc.nasa.gov/}}. We will use as a default
mask the Kp0 mask, which cuts out the Galactic plane and point sources bright enough to be resolved by WMAP,
leaving about 78.46\% of the sky \cite{Bennett:2003bz}. 

The intrinsic quality of this dataset leaves us with few instrumental systematic effects to worry
about \cite{Jarosik:2003fe,Page:2003eu,Jarosik:2006ib}. 
Nonetheless, noise inhomogeneities and beam effects could be of particular concern for our lensing statistic.
The former will be optimally handled by our estimator. Although the latter are well
controlled for the power spectrum estimation \cite{Jarosik:2006ib,Hinshaw:2006ia}, they could 
potentially affect our lensing estimator as will be discussed
below. We will show how the formalism presented in
\cite{Hinshaw:2006ia} allows us to control them in our particular
context too. Another source of systematic error might come from other
astrophysical sources, namely residual galactic foregrounds
(synchrotron, free-free and dust), residual point sources and the
signature of galaxy clusters via the Sunyaev-Zeldovich effect. These
potential contaminants will be discussed in later sections.  

\subsection{NVSS}
\label{ssec:nvss}

As a tracer of the large scale density field, we use observations resulting
from the NRAO VLA Sky Survey (NVSS) 1.4 GHz continuum
survey. This survey covers 82\% of the sky with $\delta>-40\arcdeg$
\cite{Condon:1998iy} with a source catalog containing over $1.8\times 10^6$
sources that is 50\% complete at 2.5~mJy. It is appropriate
for our purpose since most of the bright sources are AGN-powered radio
galaxies and quasars whereas the less bright ones correspond to nearby
star-forming galaxies. As a consequence, almost all the sources away
from the Galactic plane ($|b|>2\arcdeg$) are extragalactic. 

We pixelize the NVSS catalog using HEALPix \cite{Gorski:2004by} maps
with $\Nside=256$ corresponding to around $14'$~square
pixels \footnote{For more information on HEALPix
visit  http://www.eso.org/science/healpix/}. As an
extra precaution, we removed sources with a flux greater than 1 Jy
as well as a 1 degree disk around them. We also mask out pixels at low
Galactic latitude ($|b|<10\arcdeg$) and those unobserved by the survey
($\delta<-36.87\arcdeg$). We ended up with $1.29\times 10^6$ sources 
with an average density $\overline G = 159000$ gal/steradian.

\section{CMB lensing}
\label{sec:lenstheory}

Weak lensing by large scale structure remaps the CMB temperature field on the sky;
the lensed temperature $\tT(\bn)$ and unlensed temperature $T(\bn)$ are related by
\cite{BlanchardSchneider}
\be
\tT(\bn) = T(\bn + \bd(\bn))    \label{eq:ddef}
\ee
where $\bd(\bn)$ is a vector field representing the deflection angles.
To first order in perturbation theory, $\bd(\bn)$ is expected to be a pure gradient:
\be
d_a(\bn) = \nabla_a \phi(\bn)   \label{eq:gradient}
\ee
where the scalar potential $\phi$ is given by the line of sight integral:
\be
\phi(\bn) = -2 \int_0^{\chi_*} d\chi \left( \frac{\chi_*-\chi}{\chi\chi_*} \right) 
                 \Psi(\chi \bn, \eta_0 - \chi)  \label{eq:los}
\ee
where $\chi$ denotes conformal distance along the line of sight in the assumed flat cosmology, 
$\chi_*$ is the conformal distance to recombination, and $\eta_0$ is conformal time today.
The integral in Eq.~(\ref{eq:los}) receives contributions from a broad redshift
range with median around $z\sim 2$.

How can CMB lensing be detected in data?  At the power spectrum level, lensing slightly smooths the acoustic peaks 
in the temperature power spectrum $C_\ell^{TT}$ and adds power in the damping tail \cite{Seljak:1995ve}.
However, these effects are too small to be detectable in existing datasets.  
Going beyond the power spectrum, the effect of CMB lensing on higher-point statistics of the CMB is
stronger and requires less instrumental sensitivity to detect \cite{Bernardeau:1996aa}.

The theory of CMB lens reconstruction \cite{Hu:2001fa,Hu:2001tn,Okamoto:2003zw,Hirata:2002jy}
provides a framework for extracting this higher-point signal which we will use throughout this paper.
One first defines a quadratic (in the CMB temperature $T$) estimator for the CMB lensing potential $\phi$.
The simplest higher-point estimator for detecting CMB lensing would be the power spectrum 
$C_\ell^{\phi\phi}$: a quadratic estimator in the reconstruction $\phi$ or a four-point estimator in $T$.

\begin{figure}
\centerline{\epsfxsize=3.2truein\epsffile[40 430 550 700]{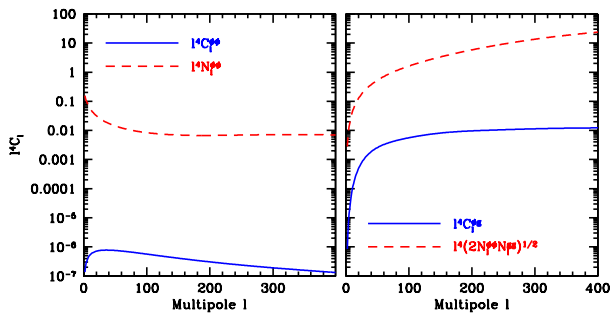}}
\caption{{\bf Left panel:} Auto power spectrum $C_\ell^{\phi\phi}$ of the CMB lensing potential, 
and reconstruction noise power spectrum $N_\ell^{\phi\phi}$ (Eq.~(\ref{eq:nlpp})) at three-year WMAP noise levels.
{\bf Right panel:} Cross power spectrum $C_\ell^{\phi g}$ between the CMB lensing potential and NVSS galaxy 
counts, and the effective noise power spectrum $[N_\ell^{\phi\phi} N_\ell^{gg} / 2]^{1/2}$
for detecting the cross-correlation.
The ``boost'' in signal-to-noise between the two cases is sufficient to obtain a several-sigma
detection of CMB lensing.}
\label{fig:reconstruction_noise}
\end{figure}

However, the three-year WMAP data do not have sufficient sensitivity to detect CMB lensing via the auto power
spectrum $C_\ell^{\phi\phi}$.
This can be seen by considering the statistical ``noise'' in the reconstruction; in \cite{Hu:2001fa} it is shown that
the reconstruction noise power spectrum $N_\ell^{\phi\phi}$ is given by:
\be
\frac{1}{N_\ell^{\phi\phi}} = \frac{1}{2\ell+1} \sum_{\ell_1\ell_2} 
   \frac{(C_{\ell_2}^{TT} F_{\ell_1\ell\ell_2} + C_{\ell_1}^{TT} F_{\ell_2\ell\ell_1})^2}
        {2(C_{\ell_1}^{TT}+N_{\ell_1}^{TT})(C_{\ell_2}^{TT}+N_{\ell_2}^{TT})}
\label{eq:nlpp}
\ee
where $F_{\ell_1\ell_2\ell_3}$ is defined by
\ba
F_{\ell_1\ell_2\ell_2} &=& \gau_{\ell_1\ell_2\ell_3} f_{\ell_1\ell_2\ell_3} \\
\gau_{\ell_1\ell_2\ell_3} &=& \sqrt{ \frac{(2\ell_1+1)(2\ell_2+1)(2\ell_3+1)}{4\pi} } 
                                \threej{\ell_1}{\ell_2}{\ell_3}{0}{0}{0}
                                \nonumber \\
&&   \\
f_{\ell_1\ell_2\ell_3} &=& \frac{\ell_2(\ell_2+1)+\ell_3(\ell_3+1)-\ell_1(\ell_1+1)}{2}  \label{eq:fdef}
\ea
In Fig.~\ref{fig:reconstruction_noise} (left panel) we have shown the noise power spectrum 
$N_\ell^{\phi\phi}$ for three-year WMAP sensitivity,
with the fiducial signal power spectrum $C_\ell^{\phi\phi}$ shown for comparison.
Although the CMB temperature anisotropies are signal-dominated across a wide range of angular scales 
(Fig.~\ref{fig:dataset_noise}), the lens reconstruction is highly noise-dominated.
At this level of signal-to-noise, an ``internal'' (to WMAP) detection of CMB lensing, by measuring the auto
power spectrum $C_\ell^{\phi\phi}$, is not possible.

It is frequently the case that a signal which is too noisy for internal detection
can nonetheless be detected via cross-correlation to a second, less noisy signal.
(For example, the first-year WMAP data had poor sensitivity to the $EE$ polarization signal,
but contained a many-sigma detection of CMB polarization via the $TE$ cross-correlation \cite{Kogut:2003et}).
In this paper, we will detect the lensing signal in WMAP by cross-correlating to radio galaxy counts in
NVSS, thus detecting a nonzero cross power spectrum $C_\ell^{\phi g}$.
The galaxy field $g$ is much less noisy than $\phi$ (Fig.~\ref{fig:dataset_noise}), but the two fields
have a significant redshift range in common and so are highly correlated;
the correlation in the fiducial model is $\sim 0.65$ on angular scales $\ell \simle 100$.
Therefore, the effective signal-to-noise is higher for the cross-correlation 
(Fig.~\ref{fig:reconstruction_noise}, right panel). 
A forecast based on this signal-to-noise ratio, and the assumption of simple $\fsky$ scaling,
predicts that a $\sim 3-4$ sigma detection can be made.  If the same forecast is repeated using
the parameters from \cite{Hirata:2004rp} (i.e. first-year WMAP sensitivity and Sloan LRG's over
4000 deg$^2$), we find a $\sim 1$ sigma result, in agreement with previous results.

In addition to the improved statistical errors from higher signal-to-noise, 
obtaining the detection as a cross-correlation is more robust to systematics, as we will see in
detail in \systematics.
Any source of systematic contamination which appears in either WMAP or NVSS, but not both, will not bias
our estimates for the cross power spectrum $C_\ell^{\phi g}$, since it does not correlate the two surveys.
At worst, such a contaminant can affect the statistical significance of the detection, by increasing the
error bars on each bandpower.

Our estimator for $C_\ell^{\phi g}$ will be defined by cross-correlating the quadratic reconstruction
of the lensing potential $\phi$ to the NVSS overdensity field $g$.  Thus the estimator is three-point: two-point in
the CMB temperature and one-point in the galaxy field.
The same three-point estimator can also be derived from the general theory of
bispectrum estimation \cite{Komatsu:2003iq,Creminelli:2005hu,Smith:2006ud}.

The most general three-point correlation between two CMB multipoles and one galaxy multipole which is
allowed by rotational and parity invariance is of the form:
\be
\langle a^T_{\ell_1 m_1} a^T_{\ell_2 m_2} a^g_{\ell_3 m_3} \rangle =
b_{\ell_1\ell_2\ell_3} \gau_{\ell_1\ell_2\ell_3}
\threej{\ell_1}{\ell_2}{\ell_3}{m_1}{m_2}{m_3}   \label{eq:bdef}
\ee
This equation defines the bispectrum $b_{\ell_1\ell_2\ell_3}$.
(More properly, with the $\gau_{\ell_1\ell_2\ell_3}$ prefactor included, we have defined the ``reduced bispectrum'' in
Eq.~(\ref{eq:bdef}); with this prefactor $b_{\ell_1\ell_2\ell_3}$ reduces to the flat sky bispectrum in the limit of 
large $\ell$ \cite{Komatsu:2001rj}.
Whenever we write bispectra in this paper, $\ell_1,\ell_2$ are understood to denote CMB multipoles
and $\ell_3$ denotes a galaxy multipole.

From this perspective, the CMB lensing signal simply gives a contribution to the bispectrum which we
want to measure.  The lensing bispectrum is proportional to $C_\ell^{\phi g}$:
\be
b_{\ell_1\ell_2\ell_3} = (f_{\ell_1\ell_2\ell_3} C_{\ell_2}^{TT}
      + f_{\ell_2\ell_1\ell_3} C_{\ell_1}^{TT}) C_{\ell_3}^{\phi g}   \label{eq:Blensing}
\ee
One can think of this as a single bispectrum which is estimated to give an overall detection,
or a linear combination of independent bispectra corresponding to bandpowers in $C_\ell^{\phi g}$.

In Appendix~\ref{app:est}, we show that the lens reconstruction and bispectrum formalisms are equivalent,
so that it is a matter of convenience which to use.
In this paper, we have generally used the lens reconstruction formalism, but will occasionally
refer to the bispectrum formalism when it provides additional perspective.

\begin{figure}
\centerline{\epsfxsize=3.2truein\epsffile[90 490 550 700]{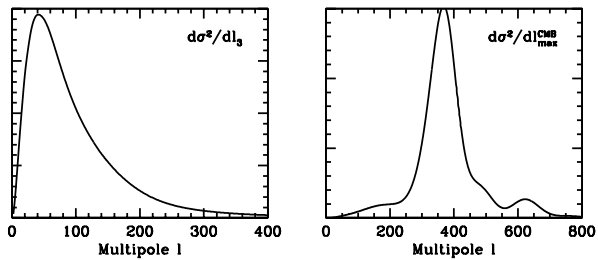}}
\caption{Mean contribution to the squared total detection significance $\sigma^2$ per NVSS galaxy multipole $\ell_3$
(left panel), and per unit increase in maximum CMB multipole $\ell_{\rm max}^{\rm CMB} = \mbox{max}(\ell_1,\ell_2)$
(right panel).
Most of the statistical weight comes from galaxy multipoles near $\ell\sim 50$, and CMB multipoles near
$\ell\sim 400$.}
\label{fig:statistical_weight}
\end{figure}

One issue which is clearer from the bispectrum perspective is the distribution of statistical weight.
Suppose we consider the total squared detection significance $\sigma^2$, rather than splitting the signal into
bandpowers.  Starting from the bispectrum in Eq.~(\ref{eq:Blensing}), one can write $\sigma^2$
as a sum over multipoles $(\ell_1,\ell_2,\ell_3)$.
In Fig.~\ref{fig:statistical_weight}, we have split up this sum to show the contribution per multipole.
(Since there are two CMB multipoles, we show the contribution per unit increase in the maximum
multipole $\ell_{\rm max}^{\rm CMB} = \mbox{max}(\ell_1,\ell_2)$.)
It is seen that the greatest statistical weight comes from galaxy multipoles near $\ell_3 \sim 50$, and
CMB multipoles near $\ell\sim 400$ corresponding to an acoustic trough in the primary CMB.
In bispectrum language, most of the signal is in ``squeezed'' triangles where the galaxy wavenumber is
much smaller than the two CMB wavenumbers.
This corresponds to the intuitive statement that lens reconstruction estimates degree-scale
lenses indirectly through their effect on smaller-scale hot and cold spots in the CMB.

\section{Pipeline}
\label{sec:pipeline}

\begin{figure}
\begin{center}
\begin{picture}(200,250)
\put(85,235){\scriptsize (1)}
\put(100,245){\vector(0,-1){18}}
\put(100,220){\makebox(0,0){Gaussian fields \{$g_{\ell m}$, $\phi_{\ell m}$, $a_{\ell m}^{\rm unlensed}$\}}}
\put(50,215){\vector(0,-1){18}}
\put(35,205){\scriptsize (2)}
\put(150,213){\vector(0,-1){48}}
\put(135,190){\scriptsize (4)}
\put(50,190){\makebox(0,0){Lensed CMB $a_{\ell m}^{\rm lensed}$}}
\put(50,185){\vector(0,-1){20}}
\put(35,174){\scriptsize (3)}
\put(115,120){\framebox(70,40){\epsfxsize=1.0truein\epsffile{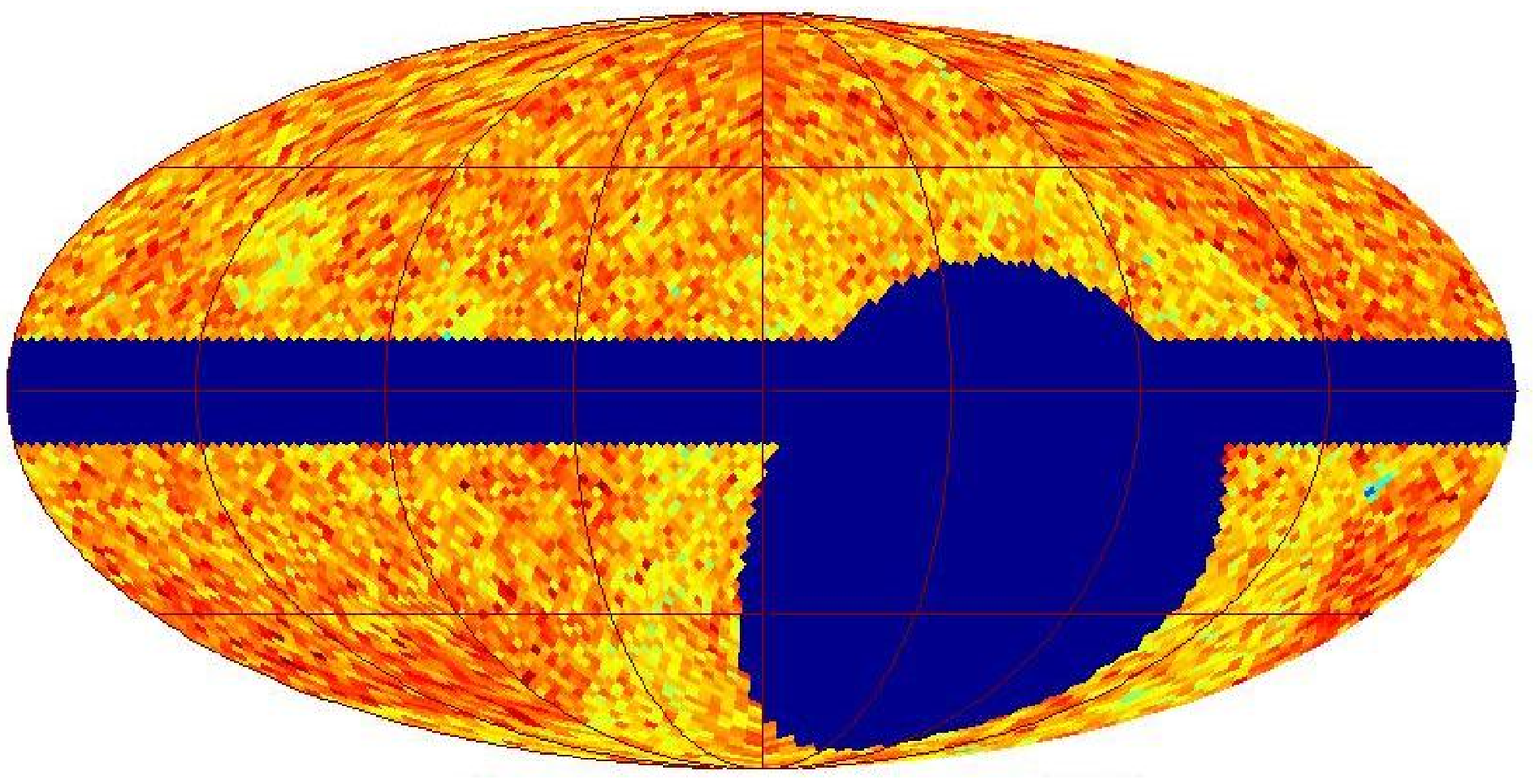}}}
\put(150,111){\makebox(0,0){NVSS data}}
\put(150,106){\vector(0,-1){28}}
\put(135,90){\scriptsize (7)}
\put(150,70){\makebox(0,0){Filtered galaxy}}
\put(150,59){\makebox(0,0){field $\tg_{\ell m}$}}
\put(132,54){\vector(-1,-2){22}}
\put(50,141){\makebox(0,0){\epsfxsize=1.0truein\epsffile{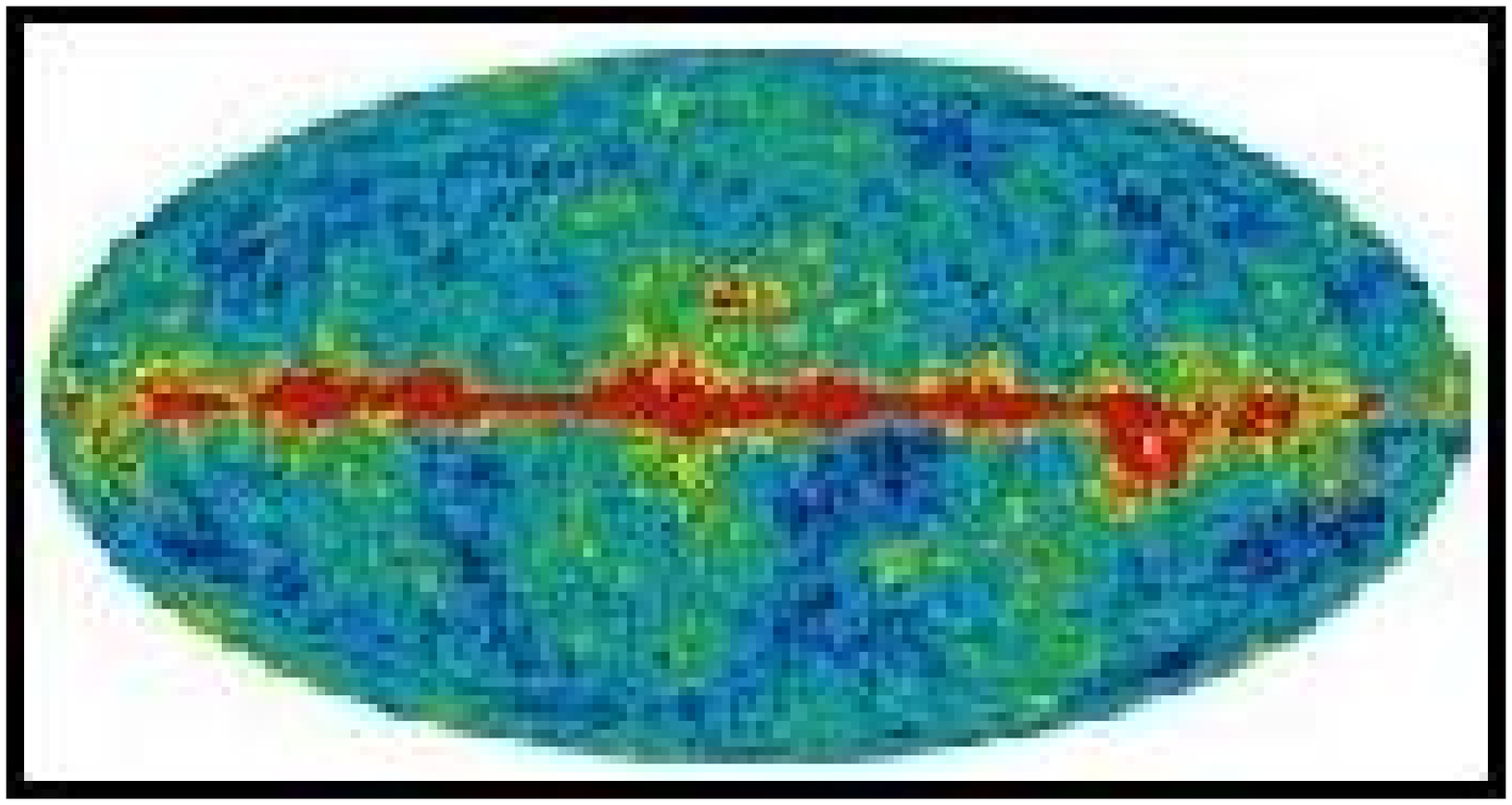}}}
\put(50,111){\makebox(0,0){WMAP data}}
\put(50,105){\vector(0,-1){18}}
\put(35,95){\scriptsize (5)}
\put(50,81){\makebox(0,0){Filtered CMB $\ta_{\ell m}$}}
\put(50,76){\vector(0,-1){18}}
\put(35,66){\scriptsize (6)}
\put(50,51){\makebox(0,0){Reconstructed}}
\put(50,40){\makebox(0,0){potential $\tphi_{\ell m}$}}
\put(80,35){\vector(1,-1){25}}
\put(100,20){\scriptsize (8)}
\put(100,5){\makebox(0,0){Lensing estimator $\hC_b^{\phi g}$}}
\end{picture}
\end{center}
\caption{Simulation + analysis pipeline used in this paper; the stages (1)-(7) are described in detail in \S\ref{sec:pipeline}.}
\label{fig:pipeline}
\end{figure}

In this section, we describe our simulation and analysis pipeline for estimating
the cross power spectrum $C_\ell^{\phi g}$ from the WMAP and NVSS datasets, and 
present results with statistical errors.  (Systematics will be treated in \systematics.)

\subsection{Pipeline description}
\label{ssec:pipeline_description}

Our pipeline is shown in Fig.~\ref{fig:pipeline}.  
Steps (1)-(4) represent the simulation direction and produce simulated WMAP and NVSS datasets with CMB lensing.
Steps (5)-(8) are the analysis direction
and produce power spectrum estimates $\hC_b^{\phi g}$ in bands $b$, starting from the WMAP and NVSS datasets.
We now describe each step in detail.

The first step (1) is simulating Gaussian fields: the unlensed CMB temperature, lensing potential,
and (shot noise free) radio galaxy field $g$.
We use the power spectra $C_\ell^{TT}, C_\ell^{\phi\phi}, C_\ell^{gg}, C_\ell^{\phi g}$ in the fiducial model.
The last two are computed using the Limber approximation  (e.g. \cite{Bartelmann:1999yn}) and a simple
constant galaxy bias model: we take the galaxy overdensity to be given by the line of sight integral
\be
\delta g(\bn) = b_g \frac{\int d\chi \frac{dN}{d\chi} \delta(\chi\bn,\eta_0-\chi)}{\int d\chi \frac{dN}{d\chi}}
\ee
using a fiducial redshift distribution $dN/d\chi$ and galaxy bias $b_g$ that will be discussed in
the next section.

In step (2), we compute the lensed CMB from the lensing potential and unlensed CMB.
The lensing operation
\be
\tT(\bn) = T(\bn + \bd(\bn))  \label{eq:lensop}
\ee
is performed directly in position space (rather than relying on an approximation to Eq.~(\ref{eq:lensop})
such as the gradient approximation).
The right-hand side of Eq.~(\ref{eq:lensop}) is evaluated using cubic interpolation on a high resolution
($\approx 0.5$ arcmin) map.

In step (3), we simulate the eight Q, V, and W-band channels of WMAP.
The maps are simulated at Healpix resolution $\Nside=1024$ and downgraded to $\Nside=512$ to minimize 
pixelization artifacts.
To simulate each map, we first convolve with the beam and pixel window in harmonic space:
\be
a_{\ell m} \rightarrow B_\ell W_\ell a_{\ell m}  \label{eq:beamconvolve}
\ee
where $B_\ell$ is the beam transfer function
(distinct for each channel) and $W_\ell$ is the pixel window function.
We then take the spherical transform and add Gaussian noise to each pixel.
The noise RMS is pixel-dependent but the noise is assumed uncorrelated between pixels.

As the last step in the simulation direction, in step (4) we simulate NVSS,
including clustering which is controlled by the Gaussian field $g$,
by generating a Poisson galaxy count in each pixel $p$ whose mean is given by
\be
\lambda(p) = \bar n (1 + g(p))
\ee
where $\bar n$ is the mean number of galaxies per pixel over the survey.
We simulate NVSS at $\Nside=1024$ and downgrade to $\Nside=256$.

Step (5) is the first step in the analysis direction: we start with the pixel-space maps
corresponding to the eight Q, V, and W-band WMAP channels, and compute a single harmonic-space map $\ta_{\ell m}$ 
representing the inverse signal + noise filtered temperature $\ta = (S+N)^{-1}a$.
This reduction step is a common ingredient in many types of optimal estimators 
\cite{Bond:1998zw,Oh:1998sr,Creminelli:2005hu,Smith:2006ud,Jewell:2002dz,Wandelt:2003uk}.
The general principle is that the filtering operation completely incorporates the sky cut and noise model,
so that optimal estimators can be constructed by simple subsequent operations directly in harmonic space.
For example, the optimal $TT$ power spectrum estimator is obtained by straightforwardly computing 
the power spectrum $C_\ell^{\ta\ta}$.

Here and throughout the body of the paper, we will defer technical details of the estimators
to Appendices~\ref{app:cinv},~\ref{app:est} and concentrate on conveying intuition.
In this case, the idea is that the $(S+N)^{-1}$ filter simply weights each mode of the data by
the inverse of its total variance, so that poorly measured modes are filtered out.
For example, the sky cut is incorporated into the noise covariance $N$ by assigning infinite noise variance
to pixels which are masked (in implementation, we use $N^{-1}$ rather than $N$ and set the relevant matrix 
entries to zero).
Data outside the sky cut is then completely filtered out: the map $\ta$ is independent of the map values in
masked pixels, and everything ``downstream'' in the analysis pipeline will be blind to the masked data.
As a similar example, we marginalize the CMB monopole and dipole modes by assigning them infinite variance.
Finally, the beam transfer functions (Eq.~(\ref{eq:beamconvolve})) are kept distinct in the filtering
operation, so that optimal frequency weighting is performed: the filtered map $\ta_{\ell m}$ 
will receive contributions from all frequencies at low $\ell$, but
will depend mainly on the highest-frequency channels (i.e., the channels with narrow beams) at high $\ell$.
The filtered map $\ta = (S+N)^{-1}a$ can also be thought of as the least-squares estimate of the signal,
given data from all channels.

In step (6), we perform lens reconstruction.  Given the filtered CMB temperature $\ta_{\ell m}$ from
step~(5), we compute the reconstructed potential $\tphi_{\ell m}$, defined by the equation:
\be
\sum_{\ell m} \tphi_{\ell m} Y_{\ell m}(x) = \nabla^a( \alpha(x) \nabla_a \beta(x) )  \label{eq:phidef}
\ee
where $\alpha$ and $\beta$ are defined by
\ba
\alpha(x) &=& \sum_{\ell m} \ta_{\ell m} Y_{\ell m}(x)    \label{eq:alphadef}  \\
\beta(x)  &=& \sum_{\ell m} C_\ell^{TT} \ta_{\ell m} Y_{\ell m}(x)   \label{eq:betadef}
\ea
As explained in \cite{Hu:2001fa}, $\tphi_{\ell m}$ is a noisy reconstruction of the CMB lensing potential
(or more precisely, the inverse noise weighted potential $N_\phi^{-1}\phi$, where $N_\phi$ is the noise
covariance of the reconstruction) which is quadratic in the CMB temperature.
Note that both $\ta$ and $\tphi$ are defined in harmonic space, but Eq.~(\ref{eq:phidef}) 
involves multiplication and derivative
operations in real space; in Appendix~\ref{app:est}, we explain in detail how $\tphi_{\ell m}$ is computed.

In step~(7), we perform inverse signal + noise filtering on the NVSS data: given pixel-space galaxy counts,
we compute the harmonic-space map $\tg_{\ell m} = (S+N)^{-1}g$ where the noise covariance $N$ represents shot noise.
This is analagous to the WMAP filtering operation in step~(5), but there is one new ingredient.
In addition to marginalizing data outside the sky cut, and the monopole and dipole, we marginalize any
mode which is independent of the angular coordinate $\varphi$ in equatorial coordinates.
(In harmonic space, this is equivalent to marginalizing modes with $m=0$.)
This is needed to remove a systematic effect in NVSS which we will discuss in detail in 
\S\ref{sec:nvss_systematics}; for now we remark in advance that all results in this paper 
include this marginalization.

Finally, in step (8), we compute the bandpower estimator $\hC_b^{\phi g}$ by cross-correlating
the fields $\tphi_{\ell m}$ and $\tg_{\ell m}$ from steps~(6) and~(7).  There is one wrinkle here:
as we show in Appendix~\ref{app:est}, to obtain the optimal estimator, 
we must include an extra term which subtracts the Monte Carlo 
average $\langle\tphi\rangle$ taken over unlensed simulations of WMAP:
\be
\hC_b^{\phi g} \eqdef \frac{1}{{\mathcal N}_b} \sum_{\substack{\ell\in b \\ -\ell\le m\le\ell}} 
   \frac{1}{\ell^2}
   (\tphi_{\ell m} - \langle \tphi_{\ell m} \rangle)^*
   (\tg_{\ell m})
\label{eq:onepoint}
\ee
where ${\mathcal N}_b$ is a normalization constant to be discussed shortly.
(We have included the factor $1/\ell^2$ since we estimate bandpowers assuming that $\ell^2C_\ell^{\phi g}$
is flat in each band.)
Note that the Monte Carlo average $\langle \tphi_{\ell m} \rangle$ vanishes for symmetry reasons
in the case of full sky coverage and isotropic noise, but sky cuts or noise inhomogeneities
will give rise to a nonzero average.
The extra term in Eq.~(\ref{eq:onepoint}) simply improves the variance of the estimator by 
subtracting the spurious cross-correlation between this average and the galaxy field $\tg$.

We determine the estimator normalization ${\mathcal N}_b$ by end-to-end Monte Carlo simulations
of the pipeline, including a nonzero $C_\ell^{\phi g}$ in the simulations for calibration.
(Strictly speaking, the normalization should be a matrix which couples bands $b\ne b'$, but we
have neglected the off-diagonal terms, which are small for our case of large sky coverage and
wide bands.)
As we will see in Appendix~\ref{app:est}, the normalization ${\mathcal N}_b$ is proportional to
a cut-sky Fisher matrix element, which must be computed by Monte Carlo unless an approximation is
made such as simple $\fsky$ scaling.
In addition, Monte Carlo simulations are also needed to compute the one-point term in 
Eq.~(\ref{eq:onepoint}).

This concludes our description of the pipeline.
We have not motivated the details in the construction of our lensing estimator $\hC_b^{\phi g}$,
but in Appendix~\ref{app:est} we show that the estimator is optimal, 
by proving that it achieves statistical lower limits on the estimator variance, so that the best
possible power spectrum uncertainties are obtained.
This justifies the combination of ingredients presented here: inverse signal + noise
filtering (steps 5 and 7), keeping the lensing potential in harmonic space (step 6), and
including the one-point term in the cross-correlation (step 8);
and shows that no further improvements are possible.

\subsection{Results}

\begin{figure}
\begin{center}
\centerline{\epsfxsize=3.2truein\epsffile[40 220 300 420]{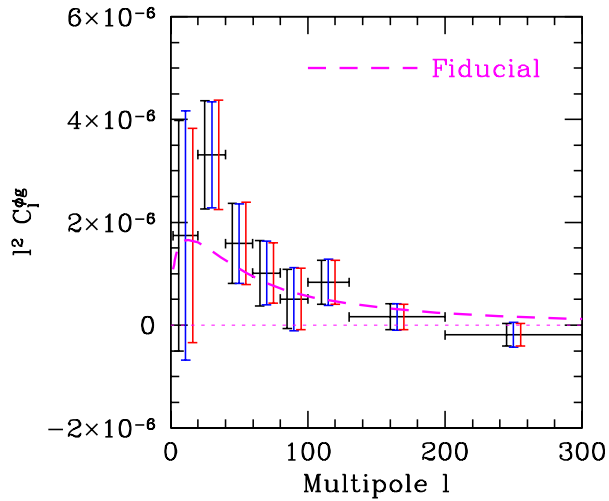}}
\end{center}
\caption{Detection of CMB lensing via the cross power spectrum $C_\ell^{\phi g}$ between
the reconstructed potential and galaxy counts.
The three 1$\sigma$ error bars on each bandpower represent different Monte Carlo methods: WMAP simulations vs NVSS 
simulations (left/black), WMAP data vs NVSS simulations (middle/blue), and WMAP simulations vs NVSS data 
(right/red). {\em These error bars represent statistical errors only; the result with systematic errors included will
be shown in Fig.~\ref{fig:final}.}}
\label{fig:statdetection}
\end{figure}

The result of applying this analysis pipeline to the WMAP and NVSS datasets is shown in 
Fig.~\ref{fig:statdetection}.
We emphasize that the uncertainties are purely statistical.
Systematic errors will be studied in \systematics, and an updated version of the result shown in \S\ref{sec:discussion},
where we also show that the detection significance with systematic errors included is 3.4$\sigma$.

Our error bars were obtained by Monte Carlo, cross-correlating simulations of WMAP and NVSS.
As a consistency check, Fig.~\ref{fig:statdetection} shows that nearly identical error bars are obtained if
WMAP simulations are cross-correlated to the real NVSS data, or vice versa.
This is an important check; if it failed, then we would know that our simulations were failing
to capture a feature of the datasets which contributes significant uncertainty to the lensing
estimator.
In addition, it shows that the uncertainties only depend on correctness of one of the simulation 
pipelines.
Suppose, for example, that the NVSS dataset contains unknown catastrophic systematics which invalidate 
our simulations.
Because the same result is obtained by treating NVSS as a black box to be cross-correlated to
WMAP simulations, it is still valid (provided that WMAP contains no ``catastrophic'' systematics!)

\begin{figure}
\begin{center}
\centerline{\epsfxsize=3.2truein\epsffile[40 220 300 420]{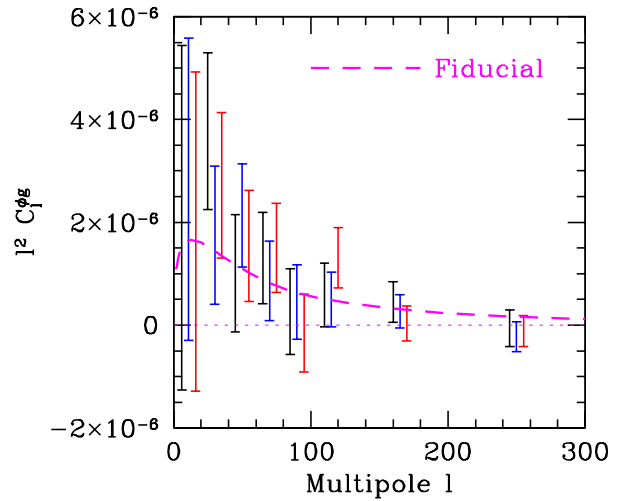}}
\end{center}
\caption{CMB lensing detection obtained by analyzing Q-band (left/black error bar in each triple),
V-band (middle/blue), and W-band (right/red) data from WMAP separately,
showing consistency of the result between CMB frequencies.}
\label{fig:freq_dependence}
\end{figure}

As another consistency check, in Fig.~\ref{fig:freq_dependence} we show the detection that is obtained if 
each frequency in WMAP is analyzed separately.
No signs of inconsistency are seen, although we have not attempted to quantify this precisely: the
results obtained from different frequencies are correlated even though the CMB noise realizations are
independent, because NVSS is identical and so is the underlying CMB realization.
For the same reason, we caution the reader that the three sets of error bars in Fig.~\ref{fig:freq_dependence}
cannot be combined in a straightforward way to obtain an overall result.
The best possible way of combining the data is already shown in Fig.~\ref{fig:statdetection}: 
the maps from the three frequencies are combined into a single CMB map which is cross-correlated to NVSS.

\subsection{Curl null test}

Our lensing estimator $\hC_b^{\phi g}$ detects a gradient component in the deflection field $d_a$
via cross-correlation to radio galaxy counts.  If we instead decompose the deflection field into 
gradient and curl:
\be
d_a(\bn) = \nabla_a \phi(\bn) + \epsilon_{ab} \nabla^b \psi(\bn)  \label{eq:gradcurl}
\ee
then one can similarly devise an estimator $\hC_b^{\psi g}$ to detect the curl component.
Since the curl component is expected to be absent cosmologically, this is a null test \cite{Cooray:2005hm}.
Note that we have parameterized the curl component by a pseudoscalar potential 
$\psi$, for notational uniformity with the gradient component which is parameterized by its scalar potential $\phi$.

In Appendix~\ref{app:est}, we show that the optimal estimator is constructed as follows.
First, we define a reconstructed potential $\tpsi$ which is quadratic in the CMB temperature:
\be
\sum_{\ell m} \tpsi_{\ell m} Y_{\ell m}(x) = \epsilon^{ab} \nabla_a( \alpha(x) \nabla_b \beta(x) )  \label{eq:psidef}
\ee
with $\alpha, \beta$ as in Eqs.~(\ref{eq:alphadef}),~(\ref{eq:betadef}).
Second, we define a power spectrum estimator by cross-correlating to galaxy counts, subtracting the one-point term:
\be
\hC_b^{\psi g} \eqdef \frac{1}{{\mathcal N}_b} \sum_{\substack{\ell\in b \\ -\ell\le m\le\ell}} 
   \frac{1}{\ell^2}
   (\tpsi_{\ell m} - \langle \tpsi_{\ell m} \rangle)^*
   (\tg_{\ell m})
\ee
This construction is identical to our construction (Eqs.~(\ref{eq:phidef}),~(\ref{eq:onepoint})))
of the lensing estimator $\hC_b^{\phi g}$, except that a $90^\circ$ rotation has been included 
(via the antisymmetric tensor $\epsilon_{ab}$) in Eq.~(\ref{eq:psidef}).

\begin{figure}
\begin{center}
\centerline{\epsfxsize=3.2truein\epsffile[40 220 300 420]{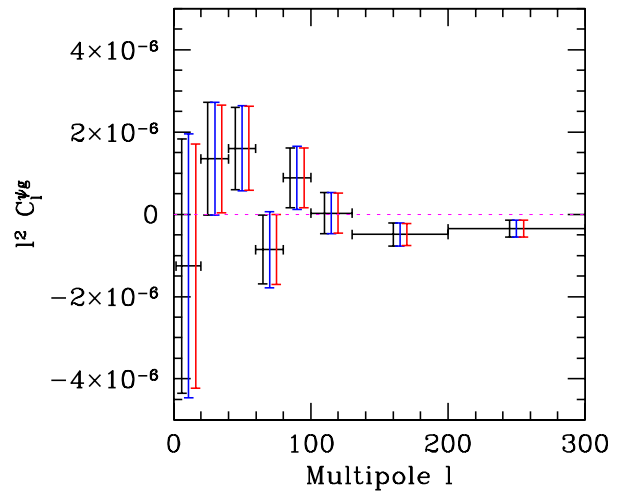}}
\end{center}
\caption{Result of the curl null test ($C_\ell^{\psi g} = 0$).  As in Fig.~\ref{fig:statdetection},
the three error bars on each bandpower represent different Monte Carlo methods: WMAP simulations vs NVSS simulations
(left/black), WMAP data vs NVSS simulations (middle/blue), and WMAP simulations vs NVSS data (right/red).}
\label{fig:nulltest}
\end{figure}

The result of the curl null test is shown in Fig.~\ref{fig:nulltest}.
The $\chi^2$ for the null test is 12.1 with 8 degrees of freedom, so the null test passes.

How strong is the null test obtained by demanding that $\hC_{\psi g}$ be consistent with zero?
One might hope that astrophysical contaminants, such as point sources or the Sunyaev-Zeldovich effect,
would contribute both gradient and curl components to the reconstructed deflections, and thus be
monitored by the null test.
However, parity invariance requires $C_\ell^{\psi g} = 0$ even when $\psi\ne 0$.  
Since astrophysical contaminants are expected to obey parity invariant statistics, they will not bias
$C_\ell^{\psi g}$ on average.
Our null test therefore only monitors contaminants which can violate parity invariance, such as Galactic
foregrounds or instrumental systematics.
This is analgous to the $C_\ell^{EB}=0$ null test in CMB polarization experiments: it is not sensitive
to all sources of contamination, but is nevertheless an important sanity check.

We remark that for a detection of CMB lensing which is internal to the CMB (detecting lensing via 
the auto power spectrum $C_\ell^{\phi\phi}$, rather than the cross spectrum $C_\ell^{\phi g}$ 
considered here), one would have one null test ($C_\ell^{\psi\psi} = 0$) which can monitor 
parity-invariant contaminants, and one null test ($C_\ell^{\phi\psi} = 0$) which cannot.

\section{NVSS systematics}
\label{sec:nvss_systematics}

In the previous section, we obtained a statistical detection of CMB lensing (Fig.~\ref{fig:statdetection})
by cross-correlating WMAP and NVSS, and showed that two consistency checks were satisfied: 
frequency independence (Fig.~\ref{fig:freq_dependence}) and a curl null test (Fig.~\ref{fig:nulltest}).
The rest of the paper is devoted to studying potential instrumental and astrophysical contaminants of 
the lensing signal, to show that the observed lensing cross-correlation is not due
to systematic contamination.
In this section, we will consider NVSS systematics.

\begin{figure}
\begin{center}
\centerline{\epsfxsize=3.2truein\epsffile[40 220 300 520]{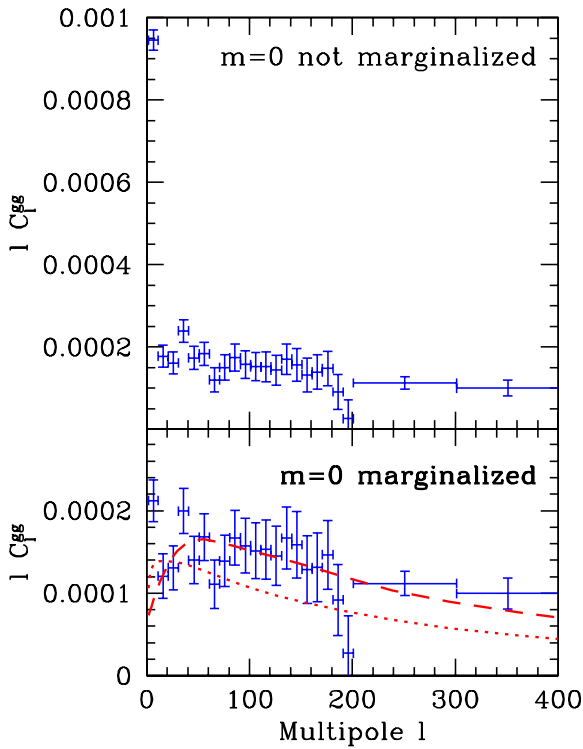}}
\end{center}
\caption{Maximum likelihood NVSS galaxy power spectrum, calculated without (top panel) and with (bottom panel) marginalization of $m=0$ modes in equatorial coordinates.  In the bottom panel, fiducial spectra are shown (both for $b_g=1.7$) from the model for $dN/dz$ by \cite{Dunlop:1990kf} (dotted line) and our fit in Eq.~\ref{eq:dnvssdz} (dashed line).
}
\label{fig:clgg}
\end{figure}

\begin{figure}
\centerline{\epsfxsize=3.2truein\epsffile{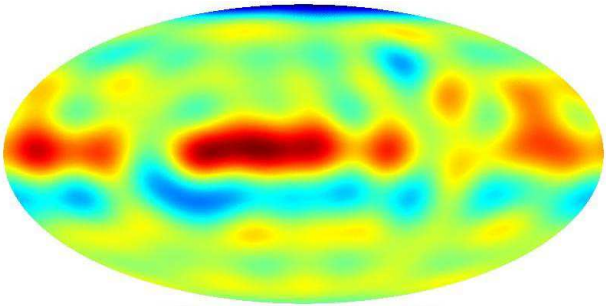}}
\caption{NVSS galaxy overdensity field in equatorial coordinates, low-pass filtered to 
multipoles $\ell\le 10$, showing visible azimuthal striping.}
\label{fig:nvss_striping}
\end{figure}

If a maximum likelihood galaxy power spectrum is calculated from NVSS
using the sky cut described in \S\ref{sec:data}, the power spectrum $C_\ell^{gg}$
shown in the top panel of Fig.~\ref{fig:clgg} is obtained.
The very high bandpower in the lowest $\ell$ band is a clear sign of systematic
contamination.
If the low $\ell$ modes are isolated by low-pass filtering the NVSS galaxy
counts to $\ell\le 10$, the resulting map shows azimuthal ``striping''
when plotted in equatorial coordinates (Fig.~\ref{fig:nvss_striping}).
This is a known systematic effect in NVSS \cite{Blake:2001bg}: due to calibration
problems at low flux densities, the galaxy density has a systematic dependence on 
declination, which can mimic long-wavelength modes in the galaxy field.

%

To remove this contaminant, we analyze NVSS in equatorial coordinates, and
marginalize any modes in the data which are constant in the azimuthal coordinate 
$\varphi$.
The marginalization is performed by modifying the NVSS noise model so that all
such modes are assigned infinite variance, as described in Appendix~\ref{app:cinv}.
Thus any signal which is constant in $\varphi$ is completely filtered out in the
inverse signal+noise weighted map $\tg$ which appears in our estimators.
Note that treating the marginalization as part of the noise model means that the
loss in sensitivity due to marginalizing $m=0$ modes is already included in the
statistical errors; it is not necessary to assign systematic errors separately.

\begin{figure}
\begin{center}
\centerline{\epsfxsize=3.2truein\epsffile[40 220 300 460]{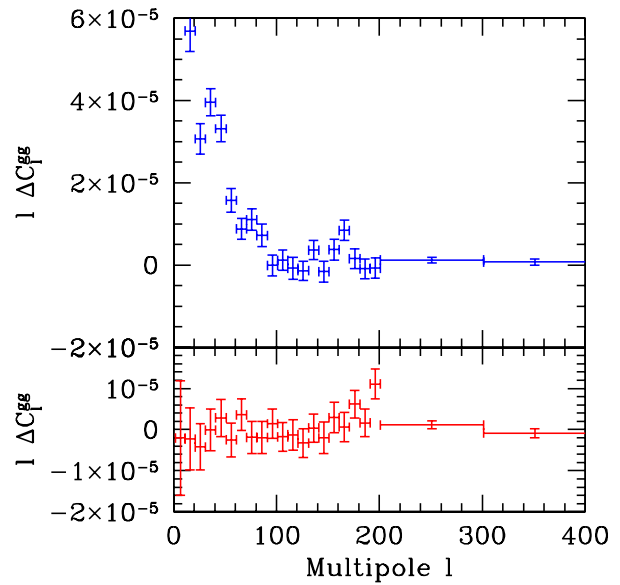}}
\end{center}
\caption{Change $\Delta C_\ell^{gg}$ in maximum likelihood galaxy power spectrum, when NVSS is
analyzed with $m=0$ marginalization vs no margnialization (top panel) or $m=0,1$ marginalization
vs $m=0$ marginalization (bottom panel) in equatorial coordinates.
The error bars represent the RMS shift obtained when Monte Carlo simulations are analyzed in the same way.}
\label{fig:deltam}
\end{figure}

After including this marginalization in the analysis, the NVSS galaxy power spectrum
shown in the bottom panel of Fig.~\ref{fig:clgg} is obtained, showing reasonable 
agreement with our fiducial $C_\ell^{gg}$.
Marginalizing $m=0$ modes produces a large shift in the lowest bandpower and
a much smaller shift in higher bands.
In Fig.~\ref{fig:deltam} (top panel), we show the shift in each bandpower when $m=0$ modes 
are marginalized, relative to an error bar which shows the RMS shift obtained when
the same marginalization is performed in NVSS simulations.
It is seen that the shift is statistically significant not only in the lowest
$\ell$ band, but all the way to $\ell\sim 100$.
We conclude that declination gradients in NVSS are an important systematic on
a range of scales and should always be marginalized in cosmological studies.

Has marginalizing $m=0$ completely removed the systematic?
To answer this, we tried marginalizing the $m=1$ Fourier mode in the azimuthal
coordinate $\varphi$, in addition to the $m=0$ mode.  
In this case, we find (Fig.~\ref{fig:deltam}, bottom panel) that the shift in $C_\ell^{gg}$
bandpowers is consistent with simulations.
(There is a possible glitch at $\ell\sim 200$, but this is outside the range
of angular scales which contribute to the lensing detection.)
Therefore, we believe that marginalizing all modes with $m=0$ in equatorial
coordinates completely removes the systematic; there is no evidence that the
contamination extends to higher $m$.

In addition to declination gradients, there is another NVSS systematic
which has been relevant for cosmological studies: multicomponent sources
\cite{Blake:2001bg,Blake:2004dg}.
Radio galaxies whose angular size is sufficiently large to be resolved by
the 45-arcsec NVSS beam will appear as multiple objects in the NVSS catalog.
This can contribute extra power to the auto spectrum $C_\ell^{gg}$, at a level 
which is a few percent of the shot noise.
At worst, this could increase the variance of our cross-correlation estimator 
$\hC_b^{\phi g}$ by a few percent without biasing the estimator.
Furthermore, as can be seen in Fig.~\ref{fig:clgg} (bottom panel), we see no evidence for 
galaxy power  in excess of fiducial in the highest $\ell$ band, which is most sensitive to 
this systematic.
We conclude that multicomponent sources are a negligible source of systematic error 
for CMB lensing.

Next we consider uncertainties in the NVSS redshift distribution $dN/dz$ and galaxy bias $b_g$.
These uncertainties affect our fiducial power spectra $C_\ell^{\phi g}, C_\ell^{gg}$ in a given
cosmology, and would need to be understood in detail if we wanted to constrain cosmological parameters
from our lensing detection.
However, since we are merely measuring the cross spectrum $C_\ell^{\phi g}$, there is only
one effect to consider: the Monte Carlo error bars we assign depend on the fiducial galaxy spectrum
$C_\ell^{gg}$ used in the simulations.
(We verified in simulations that the fiducial {\em cross} spectrum $C_\ell^{\phi g}$ does not significantly 
affect the error bars.)
If we use a fiducial $C_\ell^{gg}$ with too little power, we will underestimate our errors.
Therefore, it is important to check that our fiducial $C_\ell^{gg}$ agrees with the galaxy power
spectrum obtained from the data.

Estimates for the radio luminosity function inspired by optical and infrared observations were given in \cite{Dunlop:1990kf}. 
Using their mean-z, model 1 for average sources, \cite{Boughn:2001zs} were able to reproduce the NVSS auto-correlation function rather well. 
However the dotted curve in Fig.~\ref{fig:clgg} shows the galaxy power spectrum $C_\ell^{gg}$, 
calculated using a mean bias of $b_g=1.7$
(in agreement with the values in \cite{Boughn:2001zs,Blake:2004dg}) 
and the same model for $dN/dz$. 
For our fiducial value of $\sigma_8$, the model power spectrum is deficient relative to the observed power spectrum. 

Therefore, we search for a NVSS redshift distribution that better reproduces our angular power spectrum measurement. 
We find that for $b_g=1.7$, a near Gaussian which is lopsided toward low redshift and centered at $z_0=1.1$:
\be
\frac{dN}{dz} \propto \left\{ \begin{array}{cl}
\exp\left(-\frac{(z-z_0)^2}{2 (0.8)^2}\right) & \qquad (z < z_0) \\
\exp\left(-\frac{(z-z_0)^2}{2 (0.3)^2}\right) & \qquad (z > z_0)
\end{array} \right.  \label{eq:dnvssdz}
\ee
results in a good fit.
This match to the NVSS angular power spectrum is shown in the dashed curve of Fig.~\ref{fig:clgg}.
We have used this fiducial $C_\ell^{gg}$ in all simulations in this paper.

We make no claim that our fiducial $(dN/dz)$ is a more accurate model for the real
NVSS redshift distribution than the previously considered model.
It is just a device for generating simulations with the same power spectrum as the data, so
that we do not underestimate our error bars.
As a check, in Section \ref{sec:pipeline} we compared Monte Carlo based error estimates for WMAP 
data versus NVSS data on one hand, and WMAP data versus NVSS simulations on the other, and obtained
agreement (Fig.~\ref{fig:statdetection}).
Using the dotted line in Fig.~(\ref{fig:clgg}) would underestimate the power spectrum errors by $\sim 20\%$
due to the disagreement with the power spectrum seen in the data.
We have not investigated the reason for the disagreement in detail since it is somewhat peripheral
to the primary purpose of this paper.
However, the redshift distribution and galaxy bias assumed in the modeling would be critical 
if we were to infer constraints on cosmological parameters
(such as the normalization of matter fluctuations $\sigma_8$ or the total matter density $\Omega_0$)
from our measurements of the NVSS angular power spectrum and the cross correlation $C^{\phi g}_\ell$.
We return to this issue in \S\ref{sec:discussion}.

\section{WMAP systematics}
\label{sec:wmap_systematics}

Because our lensing estimator receives contributions from CMB anisotropies on small angular scales
(Fig.~\ref{fig:statistical_weight}), the WMAP systematics most likely to affect the detection are 
point sources and beam effects.
In our pipeline, beam effects are incorporated by convolving the CMB with an isotropic beam 
(Eq.~(\ref{eq:beamconvolve})) which is different for each DA.
This is approximate in two ways: first, the real WMAP beams are not perfectly isotropic, but contain
asymmetries which also convolve small-scale modes of the CMB by a
sky varying kernel defined by the the details of the scanning strategy. Second, the isotropic part of each beam is not
known perfectly; uncertainty in the beam transfer function acts as a source of systematic error in our lensing detection.
We study these two effects in \S\ref{ssec:beam_asymmetry},\S\ref{ssec:beam_uncertainty}.

In \S\ref{ssec:galacticforegrounds}, we consider Galactic microwave foregrounds and show that their 
effect on the lensing detection is small.  Point sources and thermal SZ will be treated separately 
in \S\ref{sec:pointsources}, \S\ref{sec:sz}.
The ISW effect \cite{Sachs:1967er} does not affect our lensing estimator, since the signal is
negligible on CMB angular scales ($\ell\sim 400$) which contribute.
The Rees-Sciama effect \cite{Rees:1968} would give a small contribution on these scales,
but we will ignore it since it is negligible compared to the SZ signal.

\subsection{Beam asymmetry}
\label{ssec:beam_asymmetry}

The WMAP beams are asymmetric due to: 1) the feeds not being at the primary focus, 
and 2) substructure caused by 0.02~cm rms
deformations in the primary mirror \cite{Page:2003eu}. 
The Q-band beams are elliptical with minor/major axis ratio of
$\approx0.8$. The V and W-band beams show significant substructure 
at the $-10$ to $-20$ dB level, leading to $\approx0.7\%$
distortions in the inferred power spectrum \cite{Hinshaw:2006ia}.

Although deviations from azimuthal symmetry of the beams have a small effect when
estimating the WMAP temperature power spectrum, it is unclear whether the same is true 
when estimating lensing.
At an intuitive level, CMB lens reconstruction recovers degree-scale modes of the lensing potential
indirectly, through their distorting effect on smaller-scale hot and cold spots in the CMB.
Beam asymmetries which convolve the small-scale CMB modes have a qualitatively similar effect and
may be degenerate with lensing.
For example, a beam quadrupole imparts an overall ellipticity or shear to the hot and cold spots.

\begin{figure}
\centerline{\epsfxsize=1.6truein\epsffile{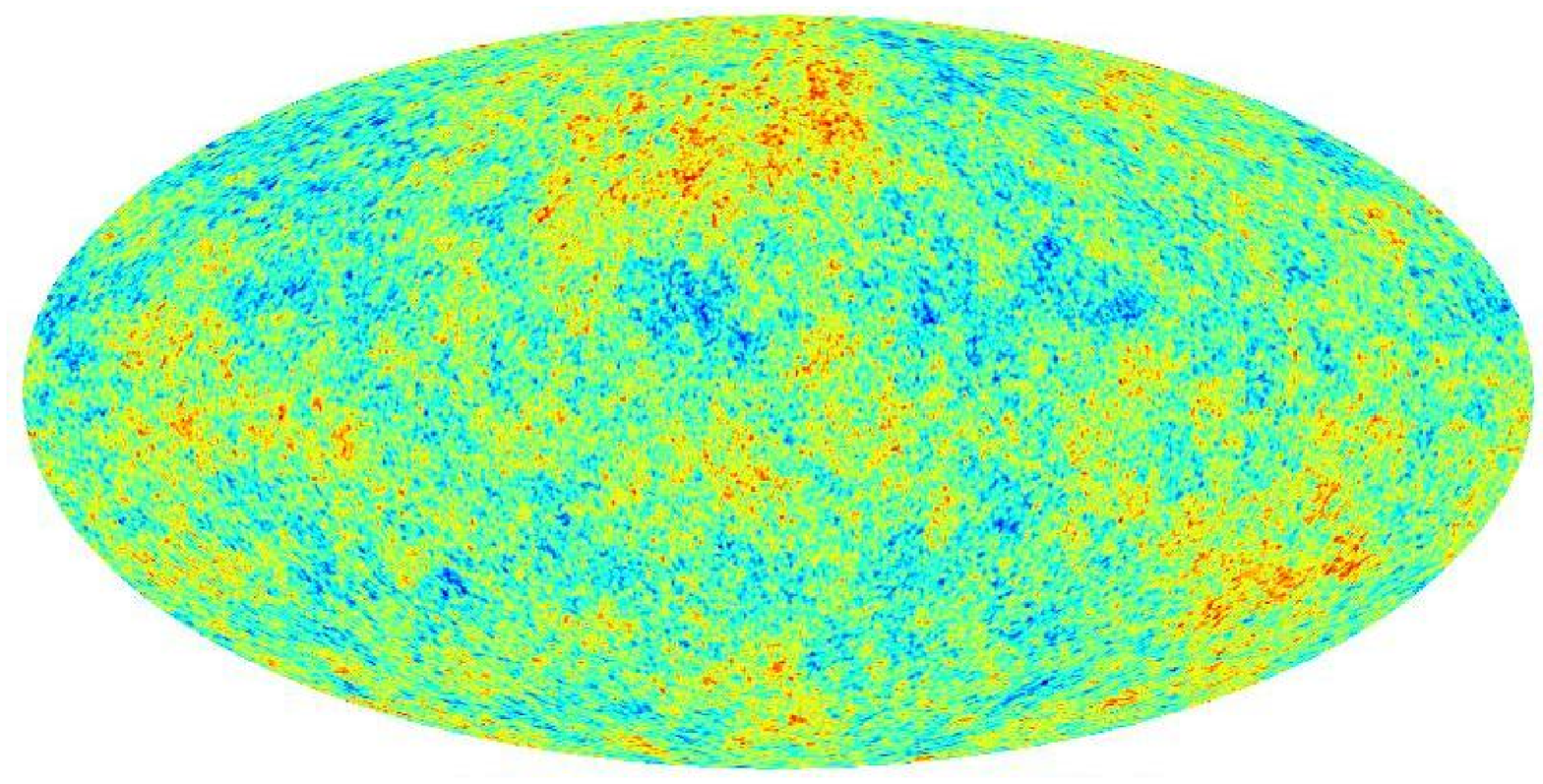} \epsfxsize=1.6truein\epsffile{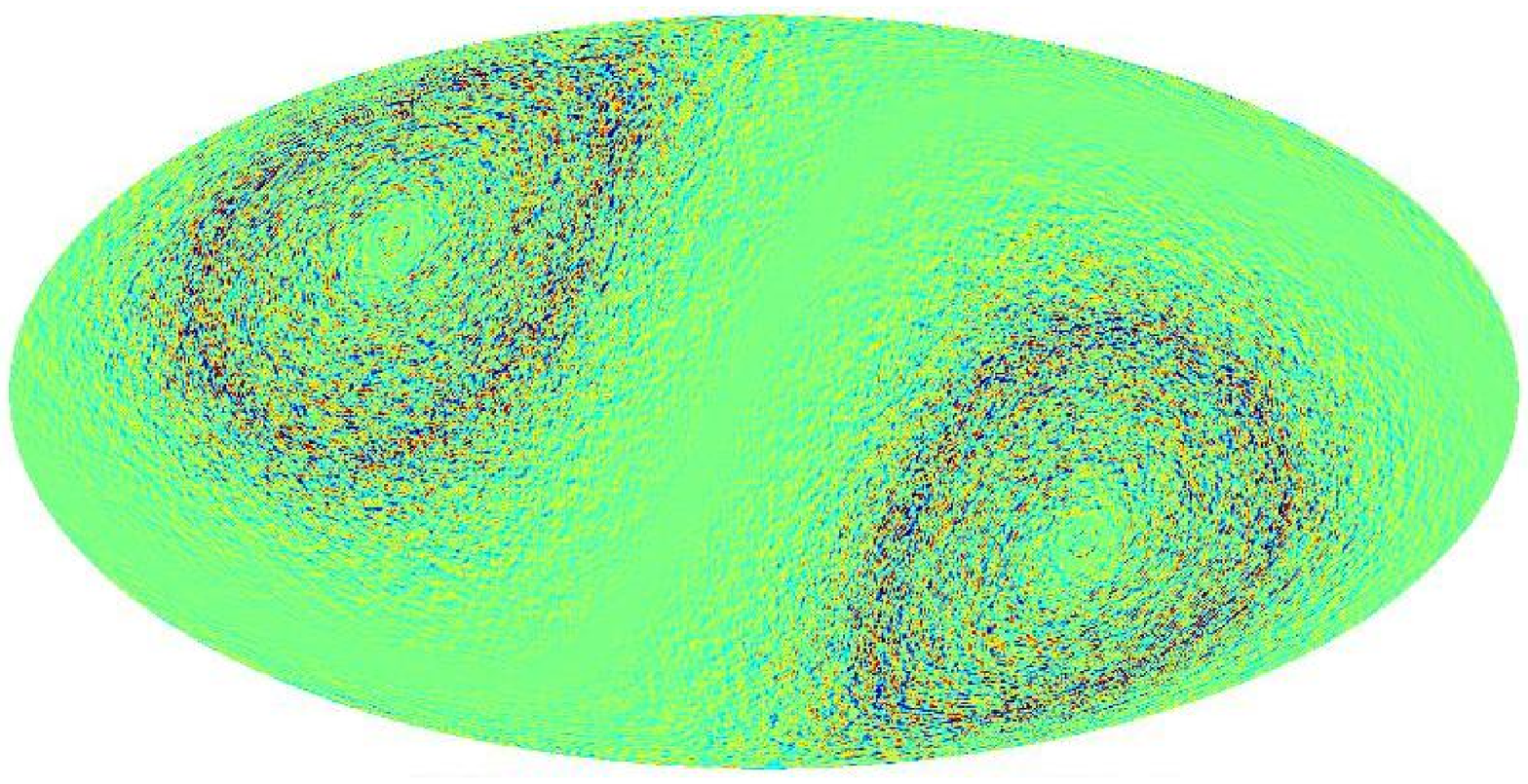}}
\centerline{\epsfxsize=1.6truein\epsffile{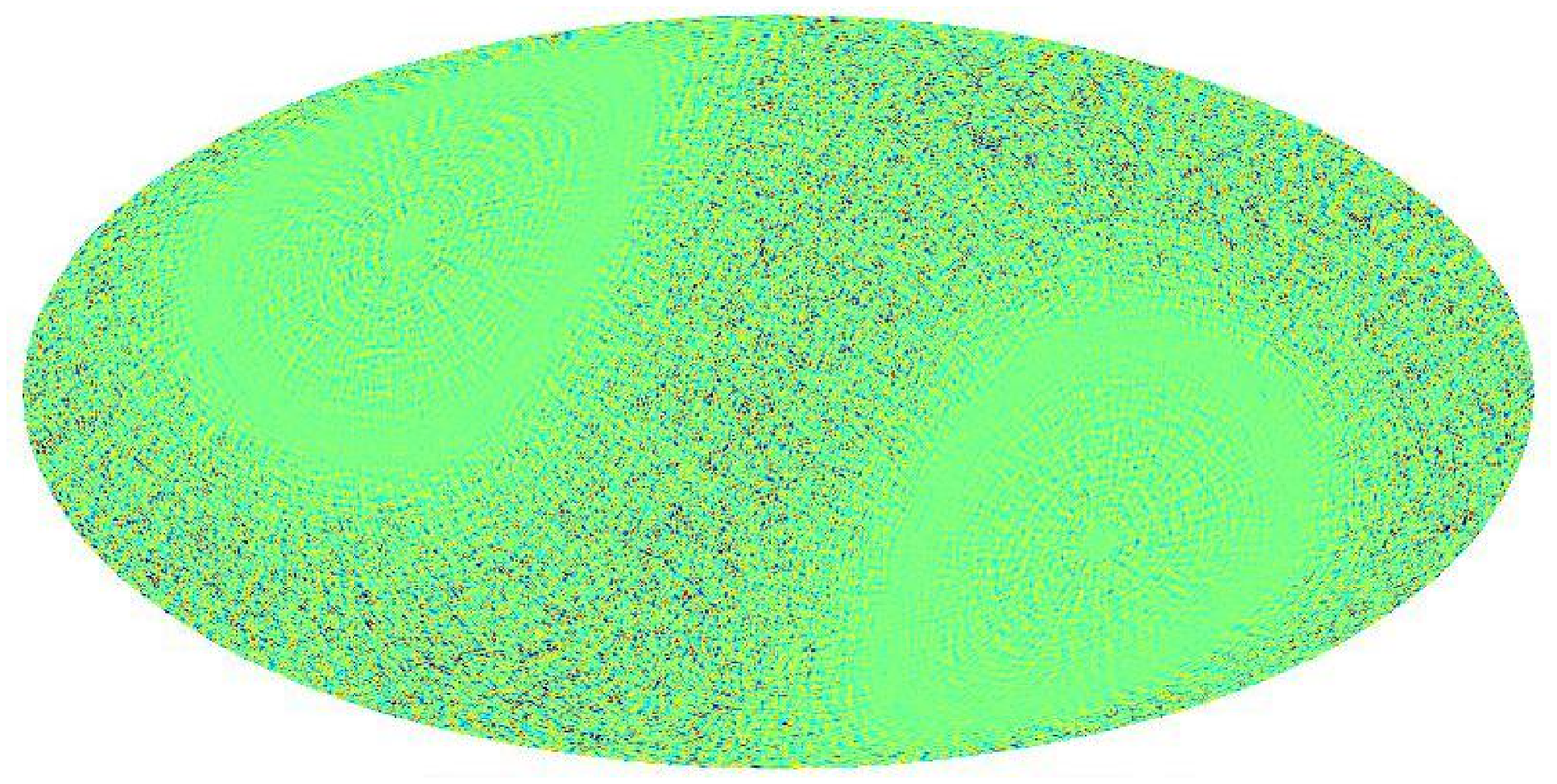} \epsfxsize=1.6truein\epsffile{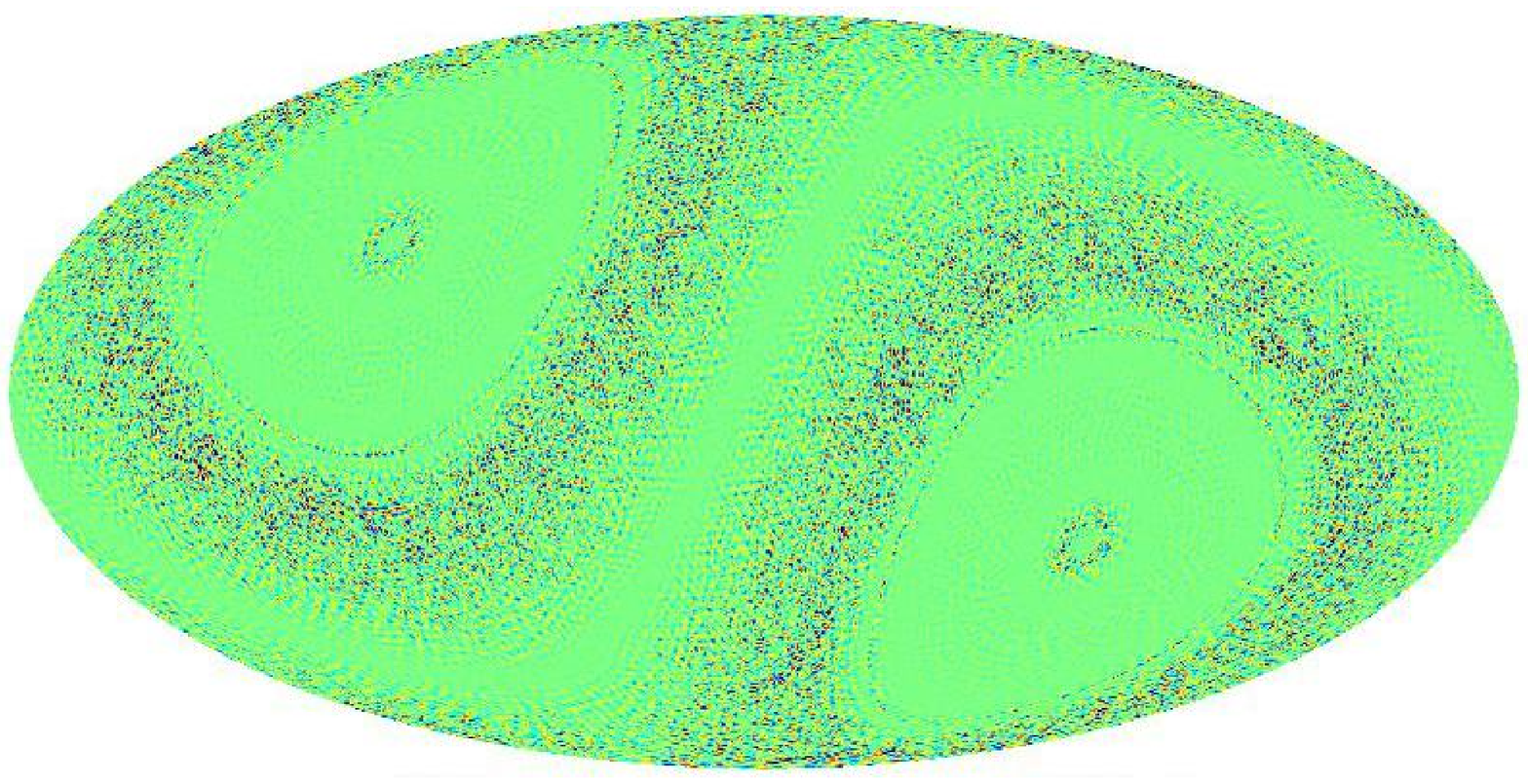}}
\caption{Result of convolving a single noiseless CMB realization with the WMAP V1 beam, including
beam asymmetry.  We have shown the output map separated into contributions from different beam
multipoles: $s=0$ (isotropic component, top left), $s=1$ (top right), $s=2$ (bottom left),
and $s=3$ (bottom right).
Each map has been scaled independently for visibility; the RMS temperature in the $s=0,\ldots,3$
maps is 88, 0.4, 1.0, 0.04 $\mu$K.
The convolution with the $s>0$ multipoles is scan dependent and shows alignments with the ecliptic
poles reflecting the WMAP scan strategy.}
\label{fig:beam_asymmetry}
\end{figure}

To incorporate beam asymmetry into our pipeline, we expand the beam profile in spherical harmonics
$Y_{\ell s}$.  The $s=0$ multipoles of the beam represent the azimuthally averaged beam and are already
incorporated in both the analysis and simulation directions of our pipeline.
The higher-$s$ multipoles have been estimated by the WMAP team and represent corrections to the
azimuthally symmetric approximation.
In Appendix~\ref{app:asymm}, we show how to incorporate the higher multipoles into the simulation
direction of the pipeline, generalizing the convolution in Eq.~(\ref{eq:beamconvolve}).
In contrast to the $s=0$ multipoles, convolving with the higher multipoles depends on the scan
strategy; our method incorporates the details of the WMAP scan based on full timestream pointing.
In Fig.~\ref{fig:beam_asymmetry}, we illustrate our simulation procedure for a single noiseless realization in
$V$-band, showing the contribution of the $s=0,\ldots 3$ multipoles to the beam-convolved map.

It would be very difficult to incorporate asymmetric beams into the analysis direction of the pipeline,
so our approach is to treat beam asymmetry as a source of systematic error.
We assign each lensing bandpower $C_b^{\phi g}$ a systematic error given by the Monte Carlo RMS {\em change} 
in the bandpower when the same WMAP + NVSS simulation is analyzed with and without including beam asymmetry
in the simulation pipeline.
We find that the systematic error in each band is small compared to the statistical error.
The result is shown, as part of a larger systematic error budget, in the ``Beam asymmetry'' column of
Tab.~\ref{tab:final} in \S\ref{sec:discussion}.

\subsection{Beam uncertainty}
\label{ssec:beam_uncertainty}

We have shown that systematic errors from beam asymmetry are small,
so that the beam may be treated as the simple
convolution in Eq.~(\ref{eq:beamconvolve}) to a good approximation.
This leaves only one remaining beam-related source of systematic error: measurement uncertainty in
the beam transfer function $B_\ell$.

We model the beam transfer function uncertainty following \cite[\S A.2]{Hinshaw:2006ia}.  The beam covariance
matrix is dominated by a small number of modes. 
We SVD decompose the matrix for each DA and keep only the 10 most significant
modes. Then we construct realizations of the beam transfer function using
\be
B_\ell = B^{(0)}_\ell\left(1+\sum_i u_i m^i_\ell \right)
\ee
where $B^{(0)}$ is the standard beam transfer function, 
$u_i$ are unit-variance normal random deviates, and $m^i_\ell$ are the
beam covariance modes.

Armed with this simulation procedure, we assign systematic errors by computing the RMS change in
each bandpower when the same simulation is analyzed with and without simulated beam uncertainty.
We find that the systematic errors are extremely small.

\subsection{Galactic foregrounds}
\label{ssec:galacticforegrounds}

\begin{figure}
\centerline{\epsfxsize=1.6truein\epsffile{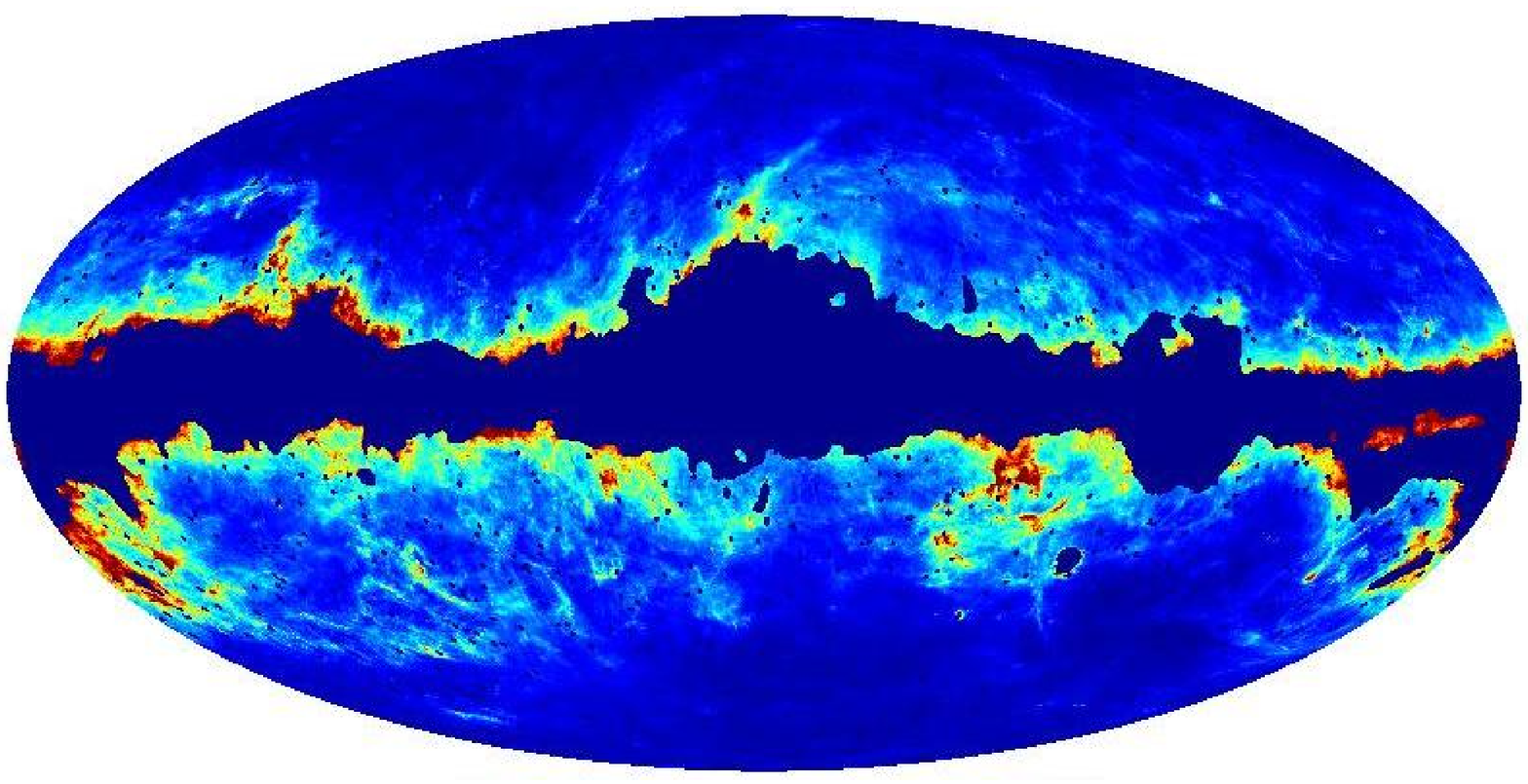} \epsfxsize=1.6truein\epsffile{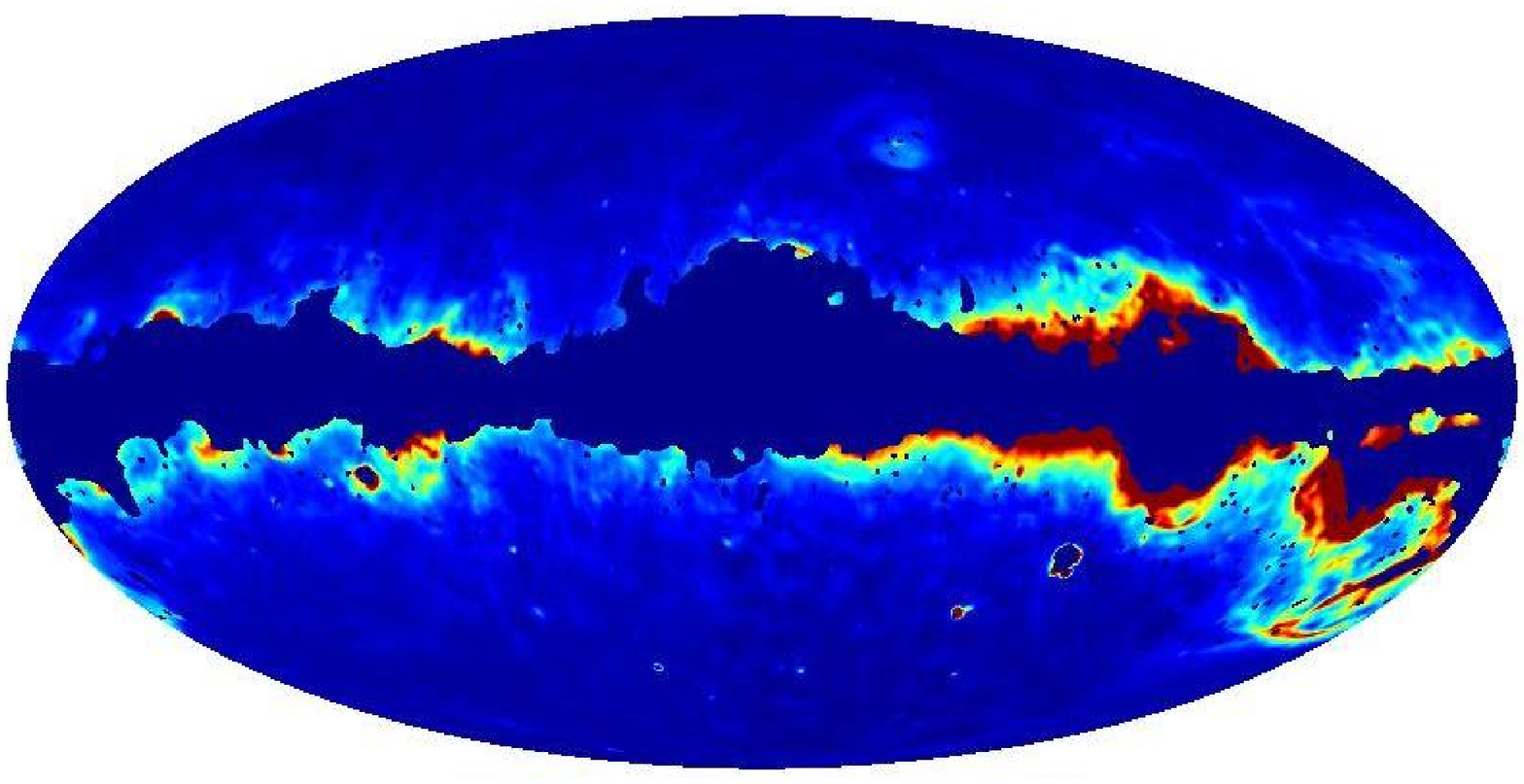}}
\caption{Foreground templates used in this paper, shown with Kp0 mask (\S\ref{sec:data}) applied.
{\bf Left panel:} Dust template, based on \cite{Finkbeiner:1999aq} with frequency dependence given by Eq.~(\ref{eq:dust_freq}).
{\bf Right panel:} Free-free template, based on \cite{Finkbeiner:2003yt,Bennett:2003ca} with frequency
dependence given by Eq.~(\ref{eq:ff_freq}).
The masked RMS of the templates in V-band is 6.4 $\mu$K and 4.8 $\mu$K respectively.}
\label{fig:fgtemplates}
\end{figure}

In addition to the CMB, the sky at microwave frequencies contains other foreground signals which
must be considered as a source of systematic error in lensing.  
We will find that the most important of these are point sources and the thermal
Sunyaev-Zeldovich effect, which will be discussed in \S\ref{sec:pointsources} and \S\ref{sec:sz}
respectively.
The other relevant microwave foregrounds are Galactic in origin: dust, free-free emission, and synchrotron 
radiation.  For descriptions of the foreground components, we refer the reader to \cite{Bennett:2003ca}.

Following \cite{Hinshaw:2006ia}, we will model dust contamination by adding a template derived from
``Model 8'' from Finkbeiner et al \cite{Finkbeiner:1999aq}, evaluated at 94 GHz and scaling to 
frequency $\nu$ by:
\be
T_A(\nu) = \left( \frac{\nu}{\mbox{94 GHz}} \right)^{2.0} T_A(\mbox{94 GHz})    \label{eq:dust_freq}
\ee
where $T_A$ denotes antenna temperature.  
The dust template is shown in Fig.~\ref{fig:fgtemplates}, left panel.

When we cross-correlate simulations of WMAP and NVSS, we find that including the dust template
in the WMAP simulation results in a very small change in the estimated lensing signal.
We take the Monte Carlo RMS average of the {\em change} in each bandpower when the same pair
of simulations is analyzed with and without the template as a systematic error estimate,
shown in the ``Dust'' column of Tab.~\ref{tab:final} in \S\ref{sec:discussion}.

One might worry that this way of assigning systematic errors, based entirely on simulations, is too
optimistic because it fails to account for unknown correlations between the templates and the
real datasets.
As a check, we obtain consistent results if we cross-correlate an ensemble of WMAP simulations against
the real NVSS data, or the real WMAP data (with and without template {\em subtraction}) against an
ensemble of NVSS simulations.
Finally, when the real WMAP and NVSS datasets are cross-correlated with and without template subtraction,
the change in each bandpower is consistent with our systematic error estimates, and no evidence for an
overall bias is seen.

We treat free-free emission similarly; in this case we use the full-sky H$\alpha$ map from \cite{Finkbeiner:2003yt},
with the correction for dust extinction from \cite{Bennett:2003ca}, and frequency dependence:
\be
T_A(\nu) = b_2 \left( \frac{\nu}{\mbox{22.8 GHz}} \right)^{-2.14} I_{H\alpha}   \label{eq:ff_freq}
\ee
where $b_2 = 6.7$ $\mu$K/Rayleigh and $I_{H\alpha}$ denotes the H$\alpha$ intensity.
Again we find consistent systematic errors in the simulation-simulation, simulation-data, and data-data
cases described in the previous paragraph.
The results are shown in the ``Free-free'' column of Tab.~\ref{tab:final} in \S\ref{sec:discussion}; 
the systematic errors from free-free are slightly higher than dust, but still small.

Finally, we turn to Galactic synchrotron emission.
The WMAP team has derived synchrotron templates both from the Haslam 408 MHz survey \cite{Bennett:2003ca,Haslam:1981xx},
and internally by differencing the K and Ka band WMAP channels \cite{Hinshaw:2006ia}.
However, both of these templates are intended for use at degree scales, and do not have sufficient
resolution to measure the sychrotron signal on the angular scales ($\ell\sim 400$) which contribute 
to our lensing estimator.
Therefore, it would not be meaningful to assign systematic errors from synchrotron emission by
using either of these templates.

In the absence of a template for synchrotron, the best we can do is to make the assumption that
the synchrotron contamination at $\ell\sim 400$ is comparable to the other Galactic foregrounds.
In V-band, synchrotron, free-free, and dust emission all contaminate 
the CMB at roughly similar levels \cite{Bennett:2003ca}.
In addition, synchrotron and dust appear to have similar spatial distributions \cite[Fig.~5]{Hinshaw:2006ia},
so the dust template should give us a reasonable estimate of possible synchroton contamination.
However, a direct test of this assumption will have to await future higher-resolution measurements
of synchrotron emission.

These results and the consistency of our measurement between frequencies (Fig.~\ref{fig:freq_dependence})
lead us to conclude that our lensing detection is not contaminated by significant residual foregrounds.
However, we quantify it by assigning each lensing bandpower a total
systematic error from foregrounds by adding the systematic errors from
the dust and free-free templates (treating the two as correlated) and
then doubling each RMS error to account for a synchrotron contribution
with the same order of  magnitude.  The result is shown in the ``Total
Galactic'' column of Tab.~\ref{tab:final} in \S\ref{sec:discussion}.

\section{Point source contamination}
\label{sec:pointsources}

Point sources which are bright enough to be resolved by WMAP are excluded by the Kp0 mask
(\S\ref{sec:data}), but unresolved point sources act as a contaminating signal in the CMB.
If the unresolved CMB point source signal were uncorrelated to NVSS, we would not expect point
sources to affect our lensing estimator $\hC^{\phi g}_b$ significantly.
However, NVSS radio galaxies will contribute some nonzero flux at microwave frequencies and
so appear directly as part of the point source contribution to the CMB.
In addition, CMB point sources which do not actually appear as objects in NVSS may be correlated
to NVSS objects in some way, e.g. if both are tracers of the same large-scale potential.
Therefore, point sources are a possible contaminant of our lensing detection.

In this section, we will place limits on the level of point source contamination and assign
systematic errors.  Point sources will turn out to be our dominant source of systematic error, 
and so we will devote considerable effort to constructing reliable error estimates.

\subsection{Point source estimator}
\label{ssec:psestimator}

It is difficult if not impossible to construct a realistic model which would allow
the level of point source contamination to be reliably estimated from general principles.
At radio and microwave frequencies, several populations of point sources have been identified
\cite{Toffolatti:1997dk,Lin:2006iu,Coble:2006ff,Giommi:2007yb} 
with significant uncertainties in spectral index and clustering properties.

Therefore, our approach will be to estimate the level of point source contamination directly
from the data.  In this subsection, we will motivate and construct an estimator which is
optimized for detecting point sources instead of CMB lensing, to use as a monitor for point
source contamination.
The first candidate for the point source estimator is simply the cross power spectrum $C_\ell^{Tg}$.

However, consider the following toy model for point sources: suppose that there are $N$
distinct populations of unclustered Poisson point sources which appear as objects in the NVSS catalog, 
and the $i$-th population has number density $n_i$ and constant flux per source $S_i$ at 
CMB frequencies.
In this model, the cross power spectrum is
\be
C_\ell^{Tg} \propto \sum_{i=1}^N S_i n_i   \label{eq:toymodel1}
\ee
whereas the bias to the lensing estimator is proportional to
\be
\DC_\ell^{\phi g} \propto \sum_{i=1}^N S_i^2 n_i  \label{eq:toymodel2}
\ee
Because the right-hand sides of Eqs.~(\ref{eq:toymodel1}),~(\ref{eq:toymodel2}) are not related in any 
model-independent way, one cannot translate a value of the cross spectrum $C_\ell^{Tg}$ to an estimate 
of the point source contamination in the lensing estimator, without making implicit assumptions about 
the point source model.

For this reason, we next consider a different candidate for the point source estimator:
the three-point estimator optimized to detect the ``Poisson'' bispectrum
\be
b_{\ell_1\ell_2\ell_3} = \mbox{constant}  \label{eq:bpoisson}
\ee
where, following Eq.~(\ref{eq:bdef}), $\ell_1,\ell_2$ denote CMB multipoles and $\ell_3$ 
denotes a galaxy multipole.  (We will construct the estimator shortly; for now we
``define'' the point source estimator by writing down the bispectrum which we want to detect.)

To motivate this form, we note that the bispectrum in our toy model is 
\be
b_{\ell_1\ell_2\ell_3} \propto \sum_{i=1}^N S_i^2 n_i
\ee
Comparing to Eq.~(\ref{eq:toymodel2}), we see that each point source population makes contributions to
the Poisson bispectrum and lensing estimator $\hC_b^{\phi g}$ which are proportional.
Therefore, an estimate of the Poisson bispectrum will directly translate to a systematic error
estimate for the lensing estimator.

This aspect of our toy model illustrates a general point:
a statistical contaminant, such as unresolved point sources, affects the lensing detection
by making a contribution to the bispectrum $b_{\ell_1\ell_2\ell_3}$ which may be coupled
to the lensing bispectrum (Eq.~(\ref{eq:Blensing})) which is measured by our estimator.
Therefore, when trying to understand point source contamination, one should first ask:
what bispectrum do point sources contribute?

We will actually consider a more general point source bispectrum than the Poisson form
in Eq.~(\ref{eq:bpoisson}), which relaxes two assumptions of the toy model.
First, we have assumed that point sources do not cluster (i.e., are purely Poisson
distributed).
Furthermore, we have assumed that each CMB point source appears as an object in NVSS;
there is a second case to consider in which the point sources do not actually appear
as objects, but are merely clustered in a way which is correlated to NVSS.

Consider a population of clustered point sources which are tracers of a Gaussian field $\rho$.
(We assume that the bias is absorbed into the definition of $\rho$, so that the probability of
a point source at position $x$ is $\propto (1+\rho(x))$.)
For our second case, where the point sources do not appear as NVSS objects, a short calculation
shows that the point source bispectrum is:
\be
b_{\ell_1\ell_2\ell_3} = \langle S^2 \rangle n C_{\ell_3}^{\rho g}  \label{eq:bps2}
\ee

In the first case, where the sources do appear as NVSS objects, the bispectrum is given by:
\be
b_{\ell_1\ell_2\ell_3} = \frac{\langle S^2\rangle n}{N} 
                           + \frac{\langle S^2\rangle n^2}{N} C_{\ell_3}^{\rho\rho}
  + \frac{\langle S \rangle^2 n^2}{N} (C_{\ell_1}^{\rho\rho} + C_{\ell_2}^{\rho\rho})  \label{eq:bps1}
\ee
where $\langle S \rangle$ is the average temperature at CMB frequencies,
$n$ is the number density of the point source population, 
and $N$ is the number density of NVSS.

In Eq.~(\ref{eq:bps1}), the first term represents contributions from Poisson statistics, the second
represents point source clustering on the galaxy angular scales ($\ell\sim 50$) which contribute 
to the lensing detection, and the third represents clustering on CMB angular scales ($\ell\sim 400$).
We will assume that the third term is small compared with the first two and can be neglected.
This is a critical assumption for our methodology and so we justify it carefully, giving
two arguments.

The first argument is that a realistic point source clustering power spectrum $C_\ell^{\rho\rho}$ 
will be rapidly decreasing with $\ell$ and so the $C_\ell$ factors in the third
term (with $\ell\sim 400$) will be small compared with the $C_\ell$ factor in the
second term (with $\ell\sim 50$).

The second argument is more formal and shows that the third term in Eq.~(\ref{eq:bps1})
is small compared to the first term.  The ratio $r$ of the third and first terms is given by
\ba
r = \frac{\langle S \rangle^2 n}{\langle S^2 \rangle} (C_{\ell_1}^{\rho\rho} + C_{\ell_2}^{\rho\rho})
      &\le& n (C_{\ell_1}^{\rho\rho} + C_{\ell_2})  \nn \\
      &\le& N (C_{\ell_1}^{gg} + C_{\ell_2}^{gg}) \nn \\
      &\simle& (1.59 \times 10^5) (2) (2.5 \times 10^{-7}) \nn \\
       &=&  0.04
\ea
In the second line, we have used the fact that the contribution to the NVSS galaxy power spectrum
$C_\ell^{gg}$ from the point source population alone is given by 
$\Delta C_\ell^{gg} = (n/N)C_\ell^{\rho\rho}$.
In the third line, we have used our measurement of $C_\ell^{gg}$ (Fig.~\ref{fig:clgg}), which shows
that $\ell C_\ell^{gg} \simle 10^{-4}$ for $\ell \simge  400$.
The intuition behind this formal argument is that if point source clustering were important on small
angular scales, we would see this signal in the NVSS power spectrum.

We have now shown that the most general point source bispectrum is a combination of 
Eq.~(\ref{eq:bps2}),
and Eq.~(\ref{eq:bps1})
with the third term neglected.
This motivates our final choice of point source estimator:
we will use the three-point estimator optimized to detect any bispectrum of the form
\be
b_{\ell_1\ell_2\ell_3} = F_{\ell_3}   \label{eq:bell3}
\ee
where $F_{\ell_3}$ is arbitrary (our estimator will estimate $F_\ell$ in bands).
This generalizes the Poisson bispectrum considered previously (Eq.~(\ref{eq:bpoisson})).

We have shown that Eq.~(\ref{eq:bell3}) is a sufficiently general form of the point source bispectrum to allow
an arbitrary clustering power spectrum between point sources, an arbitrary cross-correlation 
to the NVSS overdensity field $g$, and applies whether the CMB point sources actually
appear as objects in NVSS, or are merely correlated to NVSS.
Indeed, by putting an arbitrary $\ell_3$ dependence in Eq.~(\ref{eq:bell3}), we have been conservative
by allowing a very general point source contribution.
However, there is one caveat: we have assumed that point sources are biased 
tracers of Gaussian fields.
Non-Gaussian contributions from nonlinear evolution have not been included.
In halo model language \cite{Cooray:2002di}, we have incorporated one-halo and two-halo terms in the
bispectrum but not the three-halo term.  

Now that we have determined the most general bispectrum contributed by point source contamination
(Eq.~(\ref{eq:bell3})), how do we construct the point source estimator?
In Appendix~\ref{app:est}, we show that the optimal estimator for this bispectrum is constructed in a way which
is analagous to the lensing estimator $\hC_b^{\phi g}$ (or the curl null test $\hC_b^{\psi g}$).
First, we define a field $\ts$ which is quadratic in the CMB:
\be
\sum_{\ell m} \ts_{\ell m} Y_{\ell m}(x) = \alpha(x)^2
\ee
where $\alpha(x)$ was defined previously in Eq.~(\ref{eq:alphadef}).
Then we cross-correlate $\ts$ to galaxy counts, subtracting the one-point term as usual:
\be
\hC_b^{sg} = \frac{1}{{\mathcal N}_b} \sum_{\substack{\ell\in b \\ -\ell\le m\le\ell}} 
    (\ts_{\ell m} - \langle \ts_{\ell m} \rangle)^*  (\tg_{\ell m})  \label{eq:snull}
\ee
This defines the optimal estimator $\hC_b^{sg}$ for the point source
bispectrum (Eq.~(\ref{eq:bell3})), with the galaxy multipole $\ell_3$ binned into a bandpower $b$.

Intuitively, the field $\ts$ can be thought of as a ``quadratic reconstruction'' of CMB point source
power, in the same sense that $\tphi$ is a quadratic reconstruction of the CMB lensing potential.
Our estimator $\hC_b^{sg}$ is obtained by cross-correlating $\ts$ to the filtered galaxy field $\tg$:
we are only interested in point source power which is correlated to NVSS.
By using $\hC_b^{sg}$ to directly estimate the bispectrum due to point sources from data, 
we can assign systematic errors to the lensing bandpower $\hC_b^{\phi g}$ 
which do not depend on the details of the point source model, as we will now see.

\subsection{Results}

\begin{figure}
\begin{center}
\centerline{\epsfxsize=3.2truein\epsffile[40 220 300 420]{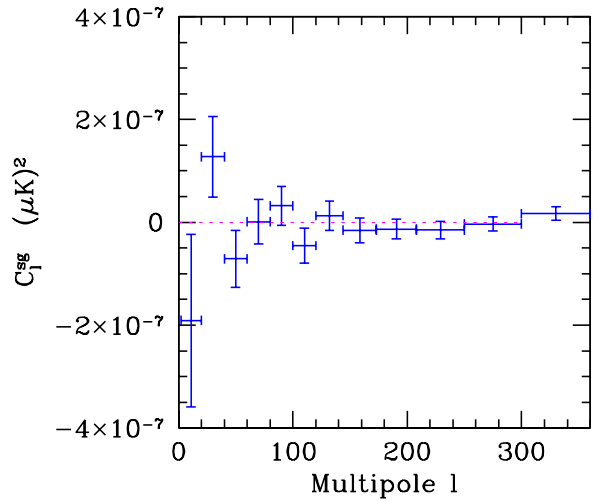}}
\end{center}
\caption{Point source estimator $\hC^{sg}_b$ applied to the WMAP and NVSS datasets, showing
no evidence for CMB point source power which is correlated to NVSS.  The error bars were obtained
from Monte Carlo WMAP+NVSS simulations without point sources.}
\label{fig:sg}
\end{figure}

In Fig.~\ref{fig:sg}, we show the result of applying the point source estimator $\hC^{sg}_b$,
constructed in the previous section, to the WMAP and NVSS datasets.  The $\chi^2$ to zero is
11.7 with 12 degrees of freedom.  Therefore, no evidence for point source contamination is seen.
This lets us put strong constraints on the systematic error in lensing due to point sources: the point 
source contribution must be small enough to be hidden in Fig.~\ref{fig:sg}, even though the
estimator $\hC^{sg}_b$ is optimized for point sources.
The rest of this subsection is devoted to assigning systematic errors based on this observation.

We find that for distinct bands $b\ne b'$, the point source and lensing estimators in band $b$
are uncorrelated to the estimators in band $b'$.
This is unsurprising; it follows from the definitions that the bands are independent for all-sky
coverage and homogeneous noise, so that the only correlation is due to inhomogeneities.
Since we have large sky coverage and wide bands, the correlations should be small.
We will treat each band independently,
for consistency with our point source model, which allows an arbitrary $\ell$
dependence in the point source amplitude (Eq.~(\ref{eq:bell3})).
We will illustrate our method in detail for the band $b = (\ellmin,\ellmax) = (20,40)$.

\begin{figure}
\centerline{\epsfxsize=3.2truein\epsffile[100 240 550 550]{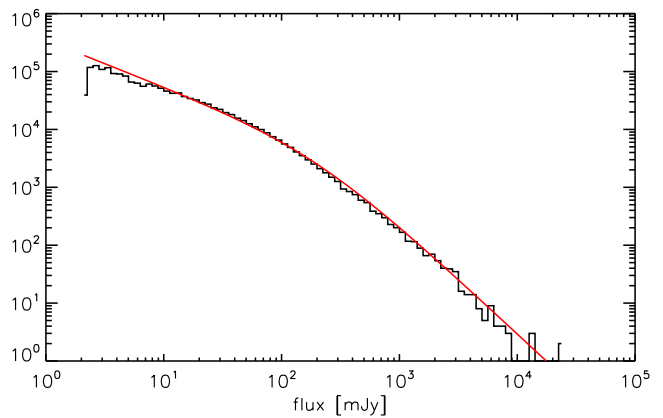}}
\caption{Histogrammed 1.4 GHz flux distribution in NVSS, with the fitting function in Eq.~(\ref{eq:mikefit})
shown for comparison.}
\label{fig:mikefit}
\end{figure}

First, we use simulations to study the effect of point sources on the estimators $\hC^{\phi g}_b,
\hC^{sg}_b$, using the following fiducial point source model.
(We will show shortly that the final result does not depend on the details of the point source model.)
Each simulated NVSS galaxy is assigned a randomly generated flux $S_\NF$
between 2 mJy and 1 Jy, drawn from the distribution
\be
\frac{dN}{dS} \propto \frac{S^{-1.8}}{1+(S/200\mbox{ mJy})^{1.1}}  \label{eq:mikefit}
\ee
This distribution was obtained empirically from the flux distribution seen in the real NVSS 
data (Fig.~\ref{fig:mikefit}).  We then assign the flux
\be
S_\nu = \Lambda \left(\frac{\nu}{1.4\mbox{ GHz}}\right)^\alpha S_\NF   \label{eq:Ldef}
\ee
at each WMAP frequency $\nu$, where $\Lambda$ is a constant which will be varied to simulate 
different overall levels of point source contamination.
Following \cite{Bennett:2003ca}, we take spectral index $\alpha=0$ in our fiducial point source model.

\begin{figure}
\begin{center}
\centerline{\epsfxsize=3.2truein\epsffile[40 220 300 420]{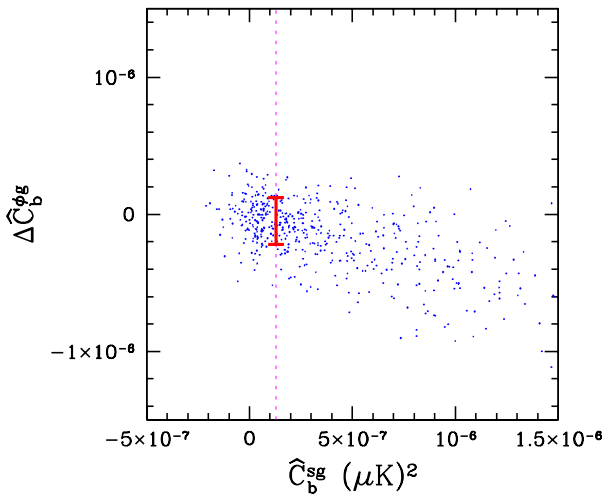}}
\end{center}
\caption{Ensemble of simulations in the fiducial point source model (Eqs.~(\ref{eq:mikefit}),~(\ref{eq:Ldef}))
with varying point source amplitude $\Lambda$.  For each realization, we show the observed point source
level $\hC_b^{sg}$ in the band $b=(\ellmin,\ellmax)=(20,40)$ and the change in the lensing estimator $\DC_b^{\phi g}$
due to the point source contribution.  The dotted vertical line shows the point source level in this band estimated from the
real WMAP + NVSS data; the smaller vertical error bar shows the mean and RMS $\DC_b^{\phi g}$ among simulations 
whose observed point source level matches the measured value.}
\label{fig:ensemble}
\end{figure}

In Fig.~\ref{fig:ensemble}, we show the values of the point source estimator $\hC^{sg}_b$ obtained
in an ensemble of simulations with varying point source amplitude $\Lambda$, and the {\em change}
$\DC^{\phi g}_b$ in the lensing bandpower which is due to the point source contribution.
(Note that we do not show the true point source amplitude $\Lambda$ for each simulation; we show the
observed point source level $\hC_b^{sg}$, estimated the same way as in the data.)

We find that the results can be fit by treating $\DC_b^{\phi g}$ as a Gaussian variable
with mean and variance which depend on $\hC_b^{sg}$:
\be
\langle \DC_b^{\phi g} \rangle = -\alpha\hC_b^{sg} \qquad \Var(\DC_b^{\phi g}) = \beta^2 + \gamma^2 (\hC_b^{sg})^2  \label{eq:ensfit}
\ee
where $\alpha=0.38$ $\mu$K$^{-2}$, $\beta=1.64\times 10^{-7}$, $\gamma=0.21$ $\mu$K$^{-2}$.

Based on this picture, how can we assign systematic errors due to point sources?
Consider the distribution of $\DC_b^{\phi g}$ values obtained by considering only realizations whose
{\em observed} point source level $\hC_b^{sg}$ agrees with the value ($=1.3\times 10^{-7}$ $\mu$K$^2$) observed 
in the data (indicated by the dotted vertical line in Fig.~\ref{fig:ensemble}.)
Note that this distribution includes realizations with a range of values for the true point source amplitude
$\Lambda$; we are effectively averaging over point source levels allowed by the 
observed value of $\hC_b^{sg}$ (i.e. the posterior distribution).
By Eq.~(\ref{eq:ensfit}), we get a Gaussian distribution with parameters:
\be
\DC_b^{\phi g} = (-0.5 \pm 1.7) \times 10^{-7}   \label{eq:ptsrcsys}
\ee
indicated by the vertical error bar in Fig.~\ref{fig:ensemble}.

We have now arrived at an distribution (Eq.~(\ref{eq:ptsrcsys})) for the change in $\DC^{\phi g}$ which is due to the 
point source contribution.  The central value of this distribution is nonzero; point source contamination makes a
negative contribution on average, as can be seen in Fig.~\ref{fig:ensemble}.
To be conservative, we will not shift our estimate for $\hC^{\phi g}$ in the positive direction by the central value 
(this would allow point sources to ``help'' the lensing detection), 
but will include the shift as part of the systematic error.
Thus we would quote the systematic error in $C_b^{\phi g}$ as: $\pm 2.2 \times 10^{-7}$.
 
As we have described it, this procedure appears to depend on the fiducial point source model
(Eqs.~(\ref{eq:mikefit}),~(\ref{eq:Ldef})).
However, we find that the final systematic error estimate in each band is relatively robust even
under drastic changes to the model.
We tried the following extreme cases: assigning constant flux to each source rather than using Eq.~(\ref{eq:mikefit}),
taking spectral index $\alpha=\pm 1$ in Eq.~(\ref{eq:Ldef}) rather than $\alpha=0$, and finally simulating point sources
which are merely correlated to NVSS rather than appearing as NVSS objects.
All of these models give similar results to within a factor $\sim 2$.
(Note that our point source estimator in Eq.~(\ref{eq:snull}) is actually optimized for point sources with a blackbody 
spectral distribution, but these results show that we obtain robust systematic error constraints across a 
reasonable range of spectral indices.)

Repeating this procedure for every $\ell$ band, we obtain a systematic error estimate for each
lensing bandpower $\hC_b^{\phi g}$.
Since we have considered several point source models, we assign the systematic error for
each band using the model which gives the largest error in that band.
The results are shown in the ``Resolved point source'' column in Tab.~\ref{tab:final} in \S\ref{sec:discussion}.
We find a systematic error which is smaller than the statistical error in all bands, but is the
largest overall source of systematic error.

The relative robustness of our error estimate to the point source model is consistent with our
discussion in the previous subsection:
regardless of the details of the model,
the contamination to the lensing estimator is proportional
to the level of the point source bispectrum (Eq.~(\ref{eq:bell3})) contributed by point sources.
By directly estimating the bispectrum, we can obtain a relatively model-independent constraint on
the systematic error due to point sources.
This would not be possible if a simpler statistic were used, such as
the cross power spectrum $C_\ell^{Tg}$.

The procedure we have described is similar to the Fisher matrix based method that is
frequently used to marginalize point sources when estimating primordial non-Gaussianity
from the CMB bispectrum \cite{Komatsu:2003iq}, but differs in several details.
First, we use a general form of the point source bispectrum (Eq.~(\ref{eq:bell3})) which
allows point source clustering, and also allows CMB point sources to appear or not appear
as NVSS objects.
Second, we do not shift the lensing estimator by the central value of the posterior distribution 
in Eq.~(\ref{eq:ptsrcsys}), but treat the shift as part of the systematic error.
Finally, the Fisher matrix formalism would not predict the increased {\em variance} in $\DC_\ell^{sg}$
in the presence of point sources (Eq.~(\ref{eq:ensfit})).
This is included in the Monte Carlo based procedure presented here.
The Fisher matrix does predict the overall negative slope in Fig.~\ref{fig:ensemble}, which
is a property of the point source and CMB lensing bispectra.
As a check, if we directly compute the Fisher matrix (see Eq.~(\ref{eq:fisheriso}) below), 
we find a weak negative correlation
($\approx -0.1$ in each band) between the lensing and point source shapes.

\subsection{Resolved point sources}
\label{ssec:mask}

\begin{figure}
\centerline{\epsfxsize=1.6truein\epsffile{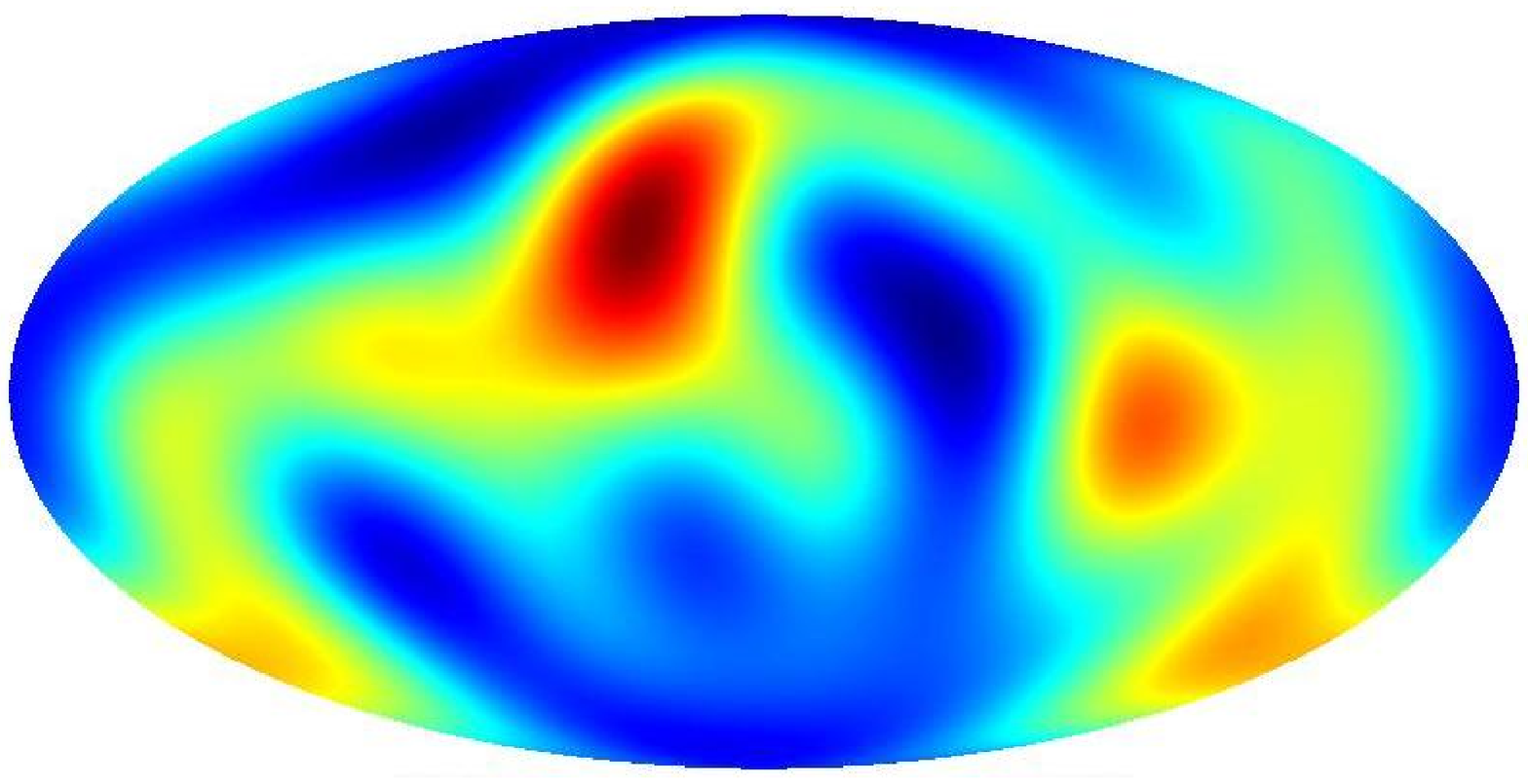} \epsfxsize=1.6truein\epsffile{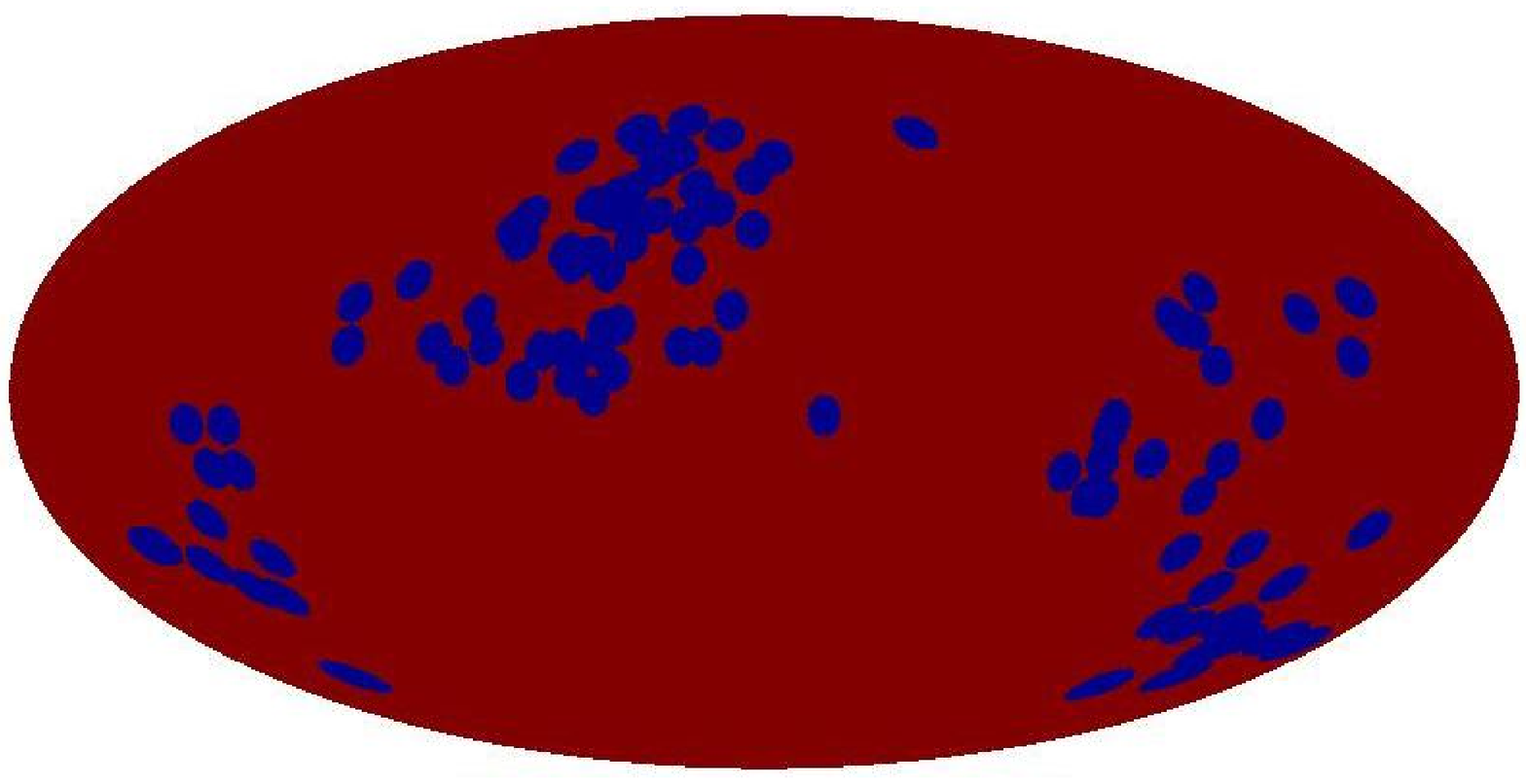}}
\caption{Illustration of our procedure (Eq.~(\ref{eq:srcmask})) for simulating correlations between
the NVSS galaxy field (left) and source mask (right).  For visibility, we have bandlimited the
galaxy field $\delta g$ to $\ell\le 6$, and used 100 sources with masking radius $4^\circ$ rather
than the mask parameters of the datasets (\S\ref{sec:data}).}
\label{fig:srcmask}
\end{figure}

Now that we have analyzed systematic errors in lensing from unresolved CMB sources,
we consider resolved sources.
Resolved CMB point sources have been treated in the pipeline by simply masking 
each source (\S\ref{sec:data}).  If the sources are correlated to radio galaxies, 
so that the WMAP {\em mask} is correlated to NVSS, one may wonder whether the
masking procedure can bias the lensing detection.

We can prove the following general result (Appendix~\ref{app:mask}): in the absence
of CMB lensing, correlations between the mask and galaxy field cannot fake the lensing
signal, i.e. the expectation value $\langle \hC_b^{\phi g} \rangle$ is
zero even if the mask is correlated.
Interestingly, our proof depends on the presence of the one-point term in the estimator
(Eq.~(\ref{eq:onepoint})) and does not rule out the possibility of bias if this term is omitted.

Given this general result, the lowest-order effect that might be expected from
mask-galaxy correlations is a bias proportional to the lensing signal, i.e. a calibration error.
We looked for a calibration error in simulations, by randomly generating a point source mask
by assigning a point source to pixel $x$ with probability
\be
\rho(x) \propto \left\{ \begin{array}{cl} \delta g(x) & \mbox{if $\delta g(x) > 0$} \\ 
                       0 & \mbox{otherwise} \end{array} \right.  \label{eq:srcmask}
\ee
(This is an extreme case, corresponding to a linear bias model $\rho(x)\propto 1 + b (\delta g(x))$
in the maximally biased limit $b\rightarrow 0$.)
An example of this simulation procedure is shown in Fig.~\ref{fig:srcmask}.

With the source mask density of the real datasets (\S\ref{sec:data}), we see no evidence for a calibration
error after 1024 Monte Carlo simulations of the full pipeline.
The same result was obtained replacing the NVSS overdensity $\delta g$ by the lensing potential $\phi$ on
the right-hand side of Eq.~(\ref{eq:srcmask}), or bandlimiting the right-hand side for
several choices of $\ell$ band.

Since we do not have a general proof that the calibration error is small, we can only conclude
that it is smaller than the $\sim 3\%$ statistical limit from our Monte Carlo sample.  In Tab.~\ref{tab:final}, 
we have assigned each bandpower a 3\% systematic calibration error in the ``Resolved point sources''
column, but we see no evidence for the effect and it may be much smaller.

\section{Sunyaev-Zeldovich fluctuations}
\label{sec:sz}

A further source of  possible contamination of the WMAP-NVSS correlation comes from 
re-scattering of the primordial microwave background off hot electrons inside the 
large scale structure field that also underlies the distribution of NVSS sources.
The largest effect 
is the thermal 
Sunyaev-Zel'dovich (SZ)  effect \cite{sz70,sz80}, due to inverse Compton scattering which 
shifts photons away from their originally black-body spectrum.  The kinetic Sunyaev-Zel'dovich (kSZ) effect, due to Doppler scattering of CMB photons by large scale structure moving along the line of sight, is expected to be a concern for lensing reconstruction with future CMB experiments that are able to frequency clean the thermal effect \cite{Amblard:2004ih,Huffenberger:2004gm}. On the angular scales relevant for WMAP, the kinetic effect is much smaller and more Gaussian than the thermal effect, and we neglect it in the following analysis.

The induced temperature change of the thermal SZ compared to the CMB, $\Delta T(\hat{n})/T_{\rm CMB}=g(\nu) y$,
is proportional to the line of sight integral over the cluster gas pressure, 
$y=\int dl n_e \frac{k_B T}{m_e c^2} \sigma_T$ (the Compton-y parameter), where $n_e$ is the free 
electron density, $k_B$ the Boltzmann constant, $T_e$ the electron temperature, $m_e$ the electron mass, 
and $\sigma_T$ the Thomson scattering cross section.
It also has a characteristic frequency dependence, given in terms of the dimensionless 
frequency $x=\frac{h \nu}{k_B T}$ by
\be
g(x)=x\frac{e^x+1}{e^x-1}-4\,.   \label{eq:szfreq}
\ee
This frequency dependence causes $\simeq 13 \, (18)\%$ changes in the  expected amplitude of the SZ between the 
WMAP V (Q) and W channels. These differences are smaller than the statistical error of our WMAP-NVSS cross 
correlation measurement, making it impossible to distinguish the SZ effect from lensing on frequency basis alone.

We therefore rely on angular separation. Our preferred way to describe the SZ effect and assign 
systematic errors would be to use full hydrodynamical simulations of the effect 
(e.g. \cite{Springel:2000bq,White:2002wp,Seljak:2001rc}). 
Unfortunately these have to date only been performed on scales of $\simeq 100$ comoving Megaparsec, 
allowing modeling of secondary anisotropies on scales of only a few square degrees. 
Our lensing estimator on the other hand receives contributions from $\ell\simge 20$, requiring simulation 
on scales substantially larger than $10\times10$ square degrees.
 A somewhat less computationally expensive route would be to establish halo catalogues based on 
perturbation theory schemes (e.g. \cite{Scoccimarro:2001cj,Monaco:2001jg}) that are then decorated with 
semi-analytic gas pressure profiles. Even these procedures  are however very costy for our purposes of 
covering 40,000 square degrees on the sky at a depth of about 4 comoving Gigaparsec, under the necessary 
requirement of resolving halos down to $10^{13}$ solar masses in order to reliably model SZ fluctuations 
below $l=1000$ \cite{Komatsu:2002wc}.

As we will argue in this section however, on the scales relevant to a lensing detection using WMAP, 
SZ contamination can be treated as part of the point source contribution which has
been studied in the previous section. 

To begin with, notice that although at WMAP frequencies SZ clusters contribute a temperature decrement to the CMB, 
their contribution to the
point source estimator $\hC_b^{sg}$ is positive, because the estimator is quadratic in the CMB.
Therefore our ``point source'' estimator will be able to serve as a monitor for the sum of point source and SZ 
contamination.
This is yet another advantage of the three-point estimator over the cross spectrum $C_\ell^{Tg}$
discussed in \S\ref{ssec:psestimator}: because point sources make a positive contribution to the cross
spectrum but the SZ contribution is negative, the cross spectrum cannot constrain both contaminants.

Next consider the spatial distribution of SZ.
The vast majority of the thermal SZ signal stems from 
collapsed regions with a gas density contrast of 
hundreds of times the mean density of the universe (see e.g. \cite{White:2002wp}).
If cluster profiles could be approximated as $\delta$-functions, then 
they could be treated as biased tracers of large scale structure that is correlated to NVSS galaxies.
Since our point source model (Eq.~(\ref{eq:bell3})) allows clustering and cross-correlation
to NVSS, this would allow us to treat the SZ contribution as part of the point source
contribution.

\begin{figure}
\includegraphics[width=6.5cm,angle=-90]{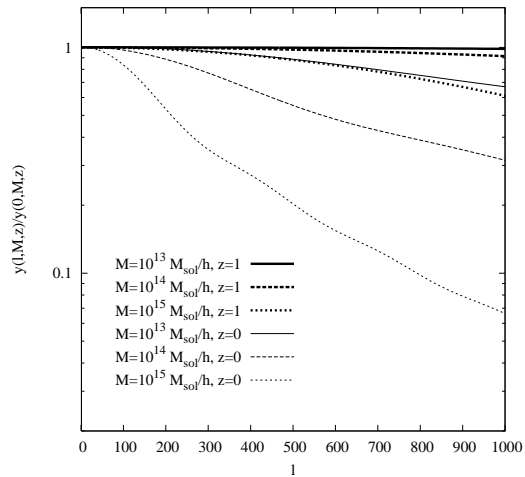}
\caption{The Compton-y profile for three different cluster masses at z=1 (thick lines) and z=0.1 (thin lines). The profiles have been normalized to 1 at l=0 to facilitate comparison. According to this panel, at high redshift it may be possible to approximate even rare and massive clusters as point sources on the scale where our lensing estimator gathers most of its information, $l \simeq 400$.}
\label{fig:comptony}
\end{figure}

To quantify the deviation from pointlike profiles, 
in Fig.~\ref{fig:comptony}, we show galaxy cluster
profiles in angular multipole space, calculated with the universal gas-pressure profile model of \cite{Komatsu:2001dn}, 
at z=0.1 and z=1.0. 
This redshift range is chosen to span roughly the range where the SZ might be correlated to NVSS 
sources. It can be seen that many of the relevant clusters fall below the angular scale 
($\ell\sim 400$) where our lensing reconstruction gathers most of its information, but 
some large nearby SZ clusters have profiles as extended as tens of arcminutes, 
and show some slope at the relevant angular scales.

\begin{figure}
\includegraphics[width=6.5cm,angle=-90]{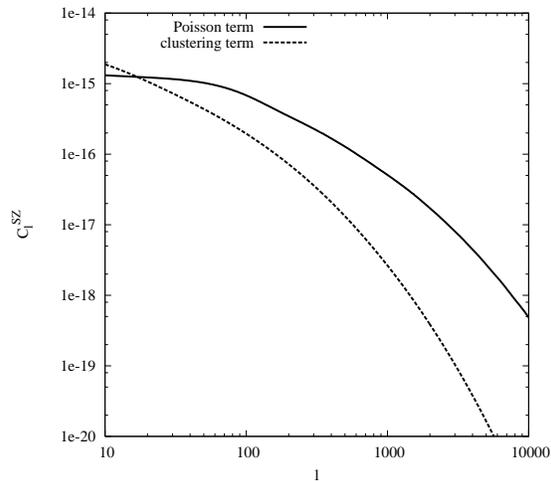}
\caption{The thermal Sunyaev-Zel'dovich angular power spectrum contributions (in the Rayleigh-Jeans limit) from Poisson and clustering terms. On the scale of most interest for our lensing reconstruction, $l\simeq 400$, the SZ Poisson term dominates by an order of magnitude over the clustering part. The angular power spectra were calculated using the gas pressure profile model by \cite{Komatsu:2001dn,Komatsu:2002wc}.}
\label{fig:clsz}
\end{figure}

To determine whether this slope is important at WMAP resolution, we consider the angular power spectrum $C_\ell^{SZ}$, which
is an average over redshift and mass of all clusters. 
In cross correlation with NVSS, this integral would be modulated by the source redshift number density. 
Since the NVSS redshift distribution is not very well understood, here we apply uniform weight to all
objects to obtain an estimate for the scale dependence of the power spectrum.
We calculate the power spectrum including both the Poisson (1-halo) and clustering (2-halo) 
contributions, following the formalism of \cite{Komatsu:1999ev,Komatsu:2002wc}. 
The results are shown (for the low frequency (Rayleigh-Jeans) limit in which $y=-2$) in Figure \ref{fig:clsz}.

It is seen that the SZ power spectrum is not flat at $\ell\sim 400$, owing to the contribution of the most massive
and nearby clusters, but has the rough scaling
\be 
C_\ell\propto \ell^{-1.2}   \label{eq:szscaling}
\ee
over the relevant range of angular scales.

We can incorporate this scale dependence into the analysis by considering a bispectrum of the form
\be
 b_{\ell_1,\ell_2,\ell_3} \propto \ell_1^{-0.6} \ell_2^{-0.6} F_{\ell_3}   \label{eq:bsz}
\ee
To quantify the effect of scale dependence on the lensing estimator, we compute the correlation between this shape and the point source shape (Eq.~(\ref{eq:bell3})),
using the Fisher matrix formalism \cite{Heavens:1998jb}.
According to this, the Fisher matrix element between two bispectra 
$b^{(\alpha)}_{\ell_1\ell_2\ell_3}, b^{(\beta)}_{\ell_1\ell_2\ell_3}$ is defined by
\be
F_{\alpha \beta} = \frac{1}{2} \sum_{\ell_1\ell_2\ell_3} \frac{(\gau_{\ell_1\ell_2\ell_3})^2 b^{(\alpha)}_{\ell_1\ell_2\ell_3} 
   b^{(\beta)}_{\ell_1\ell_2\ell_3}}{(C_{\ell_1}^{TT}+N_{\ell_1}^{TT})(C_{\ell_2}^{TT}+N_{\ell_2}^{TT})(C_{\ell_3}^{gg}+N_{\ell_3}^{gg})}  \label{eq:fisheriso}
\ee
To a good approximation, when bispectra are estimated from data, the covariance matrix is given by:
\be
\Cov(b^{(\alpha)},b^{(\beta)}) = \fsky^{-1} F^{-1}_{\alpha\beta}
\ee
When we compute the Fisher matrix for the point source (Eq.~(\ref{eq:bell3})) and scale-dependent (Eq.~(\ref{eq:bsz})) 
shapes at WMAP and NVSS noise levels, we find a correlation coefficient $\sim 0.95$.
At this level of correlation, the point source shape and SZ shape can not be distinguished to $1\sigma$, 
unless a 6$\sigma$ detection of the point source shape can also be made. 
Since we do not find any evidence for point source contamination in the data (Fig. \ref{fig:ensemble}), 
we conclude that the difference between the point source and SZ bispectra should be negligible in the context of the WMAP and NVSS data sets.

As an additional check, we tried modifying our point source simulations by giving each point source an $a_{\ell m} \propto \ell^{-0.6}$ profile,
and SZ frequency dependence (Eq.~(\ref{eq:szfreq})), including the negative sign.
This crude procedure is of course not an accurate method for simulating SZ in detail, but does incorporate two qualitative features which
distinguish SZ from point sources at WMAP resolution: the scale dependence (Eq.~\ref{eq:szscaling}) and frequency dependence (Eq.~\ref{eq:szfreq}).
We find that the systematic errors in lensing (obtained from Monte Carlo simulations as described in \S\ref{sec:pointsources}) are within the range of point source models previously considered,
showing that neither of these deviations from pure point source behavior significantly affects our method.

Finally, there is one assumption in our point source model which we can check explicitly for the
case of SZ: that clustering is unimportant on scales of $l\simeq 400$ (see Eq.~\ref{eq:bps1}).
This can be seen directly from Fig.~\ref{fig:clsz}; the clustering term is dominated by the Poisson
term by an order of magnitude.

\section{Final result and discussion}
\label{sec:discussion}

\begin{table*}
\begin{center}
\begin{tabular}{|c||c|ccc|ccc|ccc||c|}
\hline               &               &         \multicolumn{3}{c|}{Beam}     &    \multicolumn{3}{c|}{Galactic}    &   \multicolumn{3}{c||}{Point source + SZ}    &      \\
 $(\ellmin,\ellmax)$ &  Statistical  & Asymmetry  &  Uncertainty  &  Total   &  Dust & Free-free & Total &  Unresolved  &  Resolved  &  Total  &  Stat + systematic  \\   \hline
$(2,20)$    &  $17.4 \pm 22.4$  & $\pm 0.9$ & $\pm 0.3$ & $\pm 1.2$ &  $\pm 0.4$   &  $\pm 1.4$  &  $\pm 3.6$ &  $\pm 10.9$  & $\pm 0.5$ & $\pm 11.4$ & $17.4 \pm 27.4$ \\  \hline
$(20,40)$   &  $33.2 \pm 10.5$  & $\pm 0.2$ & $\pm 0.1$ & $\pm 0.3$ &  $\pm 0.2$   &  $\pm 0.5$  &  $\pm 1.4$ &  $\pm 4.9$   & $\pm 1.0$ & $\pm 5.9$  & $33.2 \pm 13.0$ \\  \hline
$(40,60)$   &  $15.9 \pm 7.8$   & $\pm 0.1$ & $\pm 0.1$ & $\pm 0.2$ &  $\pm 0.2$   &  $\pm 0.3$  &  $\pm 1.0$ &  $\pm 2.8$   & $\pm 1.5$ & $\pm 4.3$  & $15.9 \pm 9.3$ \\  \hline
$(60,80)$   &  $10.1 \pm 6.3$   & $\pm 0.1$ & $\pm 0.1$ & $\pm 0.2$ &  $\pm 0.1$   &  $\pm 0.3$  &  $\pm 0.8$ &  $\pm 2.0$   & $\pm 0.3$ & $\pm 2.3$  & $10.1 \pm 7.0$ \\  \hline
$(80,100)$  &   $5.1 \pm 5.8$   & $\pm 0.1$ & $\pm 0.1$ & $\pm 0.2$ &  $\pm 0.1$   &  $\pm 0.3$  &  $\pm 0.8$ &  $\pm 1.1$   & $\pm 0.2$ & $\pm 1.3$  & $5.1  \pm 6.0$ \\  \hline
$(100,130)$ &   $8.3 \pm 4.3$   & $\pm 0.1$ & $< 0.1$ &   $\pm 0.2$ &  $\pm 0.1$   &  $\pm 0.2$  &  $\pm 0.6$ &  $\pm 0.6$   & $\pm 0.2$ & $\pm 0.8$  & $8.3  \pm 4.4$ \\  \hline
$(130,200)$ &   $1.6 \pm 2.5$   & $< 0.1$ & $< 0.1$ &     $\pm 0.1$ &  $\pm 0.1$   &  $\pm 0.1$  &  $\pm 0.4$ &  $\pm 0.3$   & $\pm 0.1$ & $\pm 0.4$  & $1.6  \pm 2.6$ \\  \hline
$(200,300)$ &  $-1.9 \pm 2.2$   & $< 0.1$ & $< 0.1$ &     $\pm 0.1$ & $\pm 0.1$    &  $\pm 0.1$  &  $\pm 0.4$ &  $\pm 0.3$   & $\pm 0.1$ & $\pm 0.4$  & $-1.9 \pm 2.3$ \\  \hline
\end{tabular}
\end{center}
\caption{Final estimated $C^{\phi g}_b$ bandpowers, together with statistical uncertainties
and systematic errors from point sources.
All entries in the table are $\ell^2 C_\ell^{\phi g}$ in multiples of $10^{-7}$.}
\label{tab:final}
\end{table*}

\begin{figure}
\begin{center}
\centerline{\epsfxsize=3.2truein\epsffile[40 220 300 420]{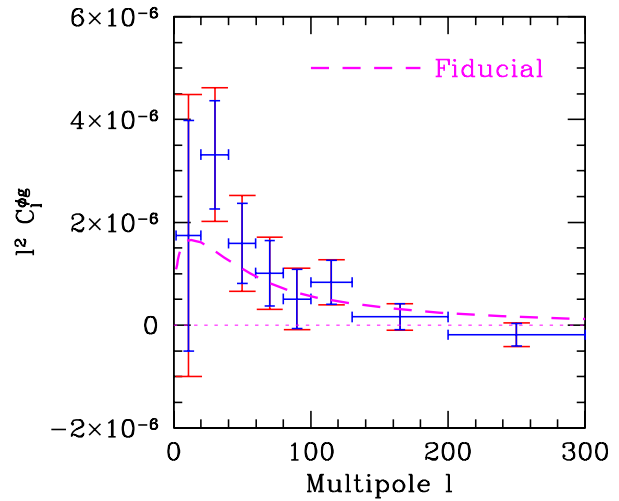}}
\end{center}
\caption{Final result from Tab.~\ref{tab:final}, showing statistical errors alone (blue/inner error bars)
and statistical + systematic errors (red/outer).}
\label{fig:final}
\end{figure}

In Tab.~\ref{tab:final} and Fig.~\ref{fig:final}, we show our final result:
an estimated value of $C^{\phi g}_b$ in bandpowers, together with statistical
and systematic uncertainties.  Our procedure for
combining errors is as follows.  We combine the errors from beam asymmetry (\S\ref{ssec:beam_asymmetry})
and beam uncertainty (\S\ref{ssec:beam_uncertainty}) into a ``total beam'' error assuming that the two are
completely correlated.  We obtain a ``total Galactic'' error from Galactic CMB foregrounds
by combining the dust and free-free systematic errors (\S\ref{ssec:galacticforegrounds}) assuming correlated 
errors, and double the result to account for synchrotron (where no template is available on the relevant 
angular scales).
We obtain a ``total point source'' error by combining the errors from unresolved and resolved sources,
assuming that the two are correlated.
(As we have shown in \S\ref{sec:sz}, the ``point source'' errors apply to the total systematic
error from CMB point sources and the thermal SZ effect.)
We then obtain our final result by 
combining the statistical, total beam, total Galactic, and point source errors, assuming that the
four are uncorrelated.

What is the total statistical significance of our detection?  To assess this, we combine our
bandpower estimates into a single estimator $\lest$, giving each bandpower a weight proportional
to its {\em fiducial} expectation value $C^{\phi g}_{b,\rm fid}$
(not the measured value in Tab.~\ref{tab:final}) and 
inversely proportional to its total (statistical + systematic) variance:
\be
\lest = \frac{\sum_b \left( C^{\phi g}_{b,\rm fid} /\Var(\hC_b^{\phi g}) \right) \hC_b^{\phi g}}
{\sum_b (C^{\phi g}_{b,\rm fid})^2 /\Var(\hC_b^{\phi g})}   \label{eq:lestdef}
\ee
where the denominator has been included to normalize $\langle\lest\rangle=1$ in the fiducial model.
We find $\lest = 1.15 \pm 0.34$, i.e. a 3.4$\sigma$ detection.

Throughout this paper, we have assumed a fiducial cosmology, NVSS redshift distribution, and galaxy bias
when computing statistical errors by Monte Carlo simulation, and when constructing the $(S+N)^{-1}$ filters
in the analysis pipeline.
To what extent do our results depend on the fiducial model?
Our $C_\ell^{\phi g}$ bandpowers and error bars depend only on the fiducial power spectra $C_\ell^{TT}, C_\ell^{gg}$
used in Monte Carlo simulations, not on the details of the modeling.
We have checked these fiducial spectra in two ways: first, by direct comparison with the measured NVSS power spectrum
(Fig.~\ref{fig:clgg}); we have omitted the comparison for the WMAP power spectrum since our fiducial cosmology is the
WMAP+ALL cosmology from \cite{Spergel:2006hy}.
Second, we have shown that consistent statistical errors are obtained by cross-correlating simulations with data
(Fig.~\ref{fig:statdetection}).
The fiducial model is also used to construct the $(S+N)^{-1}$ filtering operation, but in this case using
incorrect power spectra merely makes our estimator slightly suboptimal and does not significantly affect the
detection.

The statement that our result only depends on the fiducial spectra $C_\ell^{TT}, C_\ell^{gg}$, not on the details
of the model, would not be true if we were attempting to translate our measurement of $C_\ell^{\phi g}$ 
into a constraint on cosmological parameters.
There are several obstacles to doing so which we plan to address in future work.
First, $C_\ell^{\phi g}$ depends on cosmology but is also proportional to the NVSS galaxy bias $b_g$, 
which must be marginalized.  One possible approach is to only consider quantities such as
\be
C_\ell^{\phi g}/\sqrt{C_\ell^{gg}}    \label{eq:biasindep}
\ee
which should be independent of galaxy bias (ignoring subleties like redshift-dependent bias).
Second, the NVSS redshift distribution $dN/dz$ is uncertain and must also be marginalized over some reasonable range.
We note that the auto power spectrum $C_\ell^{gg}$, which appears in Eq.~(\ref{eq:biasindep}), is more sensitive to 
changes in $dN/dz$ than the cross spectrum $C_\ell^{\phi g}$.
A conservative approach to marginalizing over cosmological parameters as well as redshift and bias uncertainties 
would be the Markov chain Monte-Carlo (MCMC) method (compare \cite{Lewis:2002ah}) 
applied to both $C_\ell^{\phi g}$ and $C_\ell^{gg}$ constraints. 

Finally, we have not considered the impact of magnification bias: the observed NVSS galaxy field is altered
by the magnifying and demagnifying effect of gravitational lenses between the source galaxies and observer
\cite{Narayan:1989xx,Broadhurst:1994qu}.
One can think of this as adding terms to the galaxy field $g(\bn)$ which depend on the matter distribution at
intermediate redshifts along the line of sight.
This introduces additional terms in $C_\ell^{\phi g}$ which are not included in our fiducial spectrum,
and have been shown to be significant when deducing cosmological constraints from ISW measurements \cite{LoVerde:2006cj}.
In a magnified region, the galaxy surface density $g(\bn)$ receives a negative contribution (since magnification 
spreads a fixed number count over a larger area) and a postive contribution (since magnification brings new
galaxies above the flux threshhold of the survey), so the effect can have either sign.
Note that magnification bias affects the fiducial $C_\ell^{\phi g}$ in a given cosmology, but does not affect our
measured values of $C_\ell^{\phi g}$ or the statistical significance of the detection.

We have constructed an estimator for the lensing cross-correlation $C_\ell^{\phi g}$ which is probably optimal
(Appendices~\ref{app:cinv},~\ref{app:est}).  The estimator is defined in three steps.  First, we filter the
WMAP and NVSS datasets by their inverse signal + noise covariance, thus ``distilling'' the datasets to harmonic-space
maps $\ta_{\ell m}, \tg_{\ell m}$.  Second, we perform lens reconstruction on the filtered WMAP data $\ta_{\ell m}$, 
producing a noisy reconstruction $\tphi_{\ell m}$ of the CMB lensing potential which is quadratic in the data.
Third, we cross-correlate $\tphi$ and $\tg$, subtracting the one-point term.

Subtracting the one-point term is necessary to make the estimator optimal, and also eliminates systematic bias
from resolved point sources (\ref{ssec:mask}), although a systematic calibration error may remain.
Since the one-point subtraction is trivial to implement in a Monte Carlo pipeline, we recommend that it
always be used.
The other feature making our estimator optimal is full-blown $(S+N)^{-1}$ filtering (Appendix~\ref{app:cinv}).
Here, it is unclear whether the optimal filter is practically necessary; 
it may be possible 
to construct a simpler filter which approximates $(S+N)^{-1}$ and produces near-optimal 
estimates in practice.
In any case, an optimal implementation is an invaluable tool when studying candidates for such a filter, since
the results can be directly compared to optimal.

We have studied potential sources of systematic error
from known NVSS systematics (\S\ref{sec:nvss_systematics}),
WMAP beam effects (\S\ref{ssec:beam_asymmetry}-\S\ref{ssec:beam_uncertainty})
Galatic microwave foregrounds (\S\ref{ssec:galacticforegrounds}),
point sources (\S\ref{sec:pointsources}),
and the thermal Sunyaev-Zeldovich effect (\S\ref{sec:sz}).
Error estimates from each of these systematics have been included in our final result.

The most problematic systematic for CMB lensing, at least when measured in cross-correlation to large-scale 
structure, seems to be point source contamination.
In general, a statistical contaminant such as point sources affects the lensing detection by contributing
some bispectrum $b_{\ell_1\ell_2\ell_3}$ which may be correlated to the lensing bispectrum
which our estimator measures (Eq.~(\ref{eq:Blensing})).
We therefore treat point sources by directly estimating the point source bispectrum from the data, to
monitor the level of contamination and assign systematic errors.
We allow a form of the point source bispectrum (Eq.~(\ref{eq:bell3})) which is sufficiently general to include 
a wide range of point source models, including clustered sources and sources which may or may not 
appear as objects in NVSS.

We have argued that at WMAP sensitivity levels, thermal Sunyaev-Zel'dovich fluctuations due to hot gas in clusters 
of galaxies can be treated as part of the point source contribution.
We checked that the level of scale dependence in the bispectrum, introduced by large nearby objects, is
unimportant at WMAP resolution, but we do not expect this to be the case for smaller scale experiments
such as Planck \footnote{http://www.rssd.esa.int/index.php?project=Planck}, ACT \cite{Kosowsky:2004sw}, 
or SPT \cite{Ruhl:2004kv}, which will begin to observe the sky in the near future. 
In fact, even the qualitative trends we have found in Tab.~\ref{tab:final} for systematic error contributions
may be different for these future surveys, which will probe new regimes of sensitivity and resolution.
The detection from WMAP that has been presented here is a milestone toward detailed measurements of CMB 
lensing that lie ahead.

\section*{Acknowledgments}
We would like to thank Mike Nolta, who could not be listed as a coauthor according to WMAP collaboration policy, 
for key contributions to calculations and text throughout this paper.

We thank Niayesh Afshordi, Anthony Challinor, Robert Crittenden, Cora Dvorkin, Chris Hirata,
Wayne Hu, Dragan Huterer, Eiichiro Komatsu, Antony Lewis, Adam Lidz, Ue-Li Pen,
Lyman Page, David Spergel, Bruce Winstein and Matias Zaldarriaga for useful discussions.
We acknowledge use of the sunnyvale computing cluster at CITA,
and the FFTW, LAPACK, CAMB, Lenspix, and Healpix software packages,
KMS was supported by the Kavli Institure for Cosmological Physics through the grant NSF PHY-0114422. 
OZ acknowledges support by the David and Lucile Packard foundation, the Alfred P. Sloan Foundation, 
and grants NASA NNG05GJ40G and NSF AST-0506556.

\appendix

\section{Fast $(S+N)^{-1}$ filtering}
\label{app:cinv}

In this appendix, we present the details of our method for computing
the inverse signal + noise weighted map $\ta = (S+N)^{-1} a$, for either the WMAP or NVSS data.

Outside the context of lens reconstuction, this inversion problem also arises for other types of
optimal analysis in which the data is weighted by inverse signal + noise,
e.g. optimal power spectrum estimation \cite{Oh:1998sr},
power spectrum analysis by Gibbs sampling \cite{Jewell:2002dz,Wandelt:2003uk},
and bispectrum estimation \cite{Smith:2006ud}.
We expect that our method will be useful in these contexts as well.

\subsection{Conjugate gradient inversion}
\label{ssec:cg}

First, let us introduce some notation.
We assume a dataset which is specified by $\Nchan$ pixel-space maps, with a common underlying
harmonic-space signal $s_{\ell m}$.  Thus we can write
\ba
d_i^{\rm pix} = A_i s + \mbox{(noise)}   \label{eq:cgsetup}
\ea
where $A_i$ is the pointing matrix associated to the $i$-th channel.

This generality is sufficient to describe both the WMAP and NVSS datasets.
For WMAP, we have $\Nchan=8$ corresponding to the eight Q, V, and W-band differencing assemblies
used in the analysis, the signal $s_{\ell m}$ is the noiseless CMB, and each pointing matrix $A_i$
includes convolution with the pixel window function and beam of the corresponding DA.
Our convention is that the signal $s$ is defined in harmonic space, while the data $d_i^{\rm pix}$
is defined in pixel space.
Thus the operator $A_i$ in Eq.~(\ref{eq:cgsetup}) is defined by applying beam and pixel window functions 
to the harmonic-space signal (see Eq.~(\ref{eq:beamconvolve})), then taking the spherical transform to produce 
a map in pixel space.
For NVSS, we have $\Nchan=1$ corresponding to a single galaxy count map, with no beam convolution included
in the pointing matrix $A$, since the 45-arcsec NVSS beam can be neglected on angular scales ($\ell\simle 250$)
which contribute to the lensing estimator.

In \cite{Tegmark:1996qs}, it is shown that the data in Eq.~(\ref{eq:cgsetup}) 
can be reduced to a single harmonic-space map $a$,
with associated noise covariance matrix $N$, without losing information.
The map $a$ and matrix $N$ are defined by the pair of equations
\ba
N^{-1}  &=& \sum_i A_i^T (N^{\rm pix}_i)^{-1} A_i  \label{eq:ninv}  \\
N^{-1}a &=& \sum_i A_i^T (N^{\rm pix}_i)^{-1} d_i^{\rm pix}  \label{eq:ninva}
\ea
where $N^{\rm pix}_i$ is the noise covariance associated to the $i$-th map.

Let us first assume a noise model (which we will generalize in \S\ref{ssec:complexnoise})
such that the inverse noise covariance $(N^{\rm pix}_i)^{-1}$ in the $i$-th map is diagonal in pixel space.
For WMAP, this is the noise model used to analyze the temperature power spectrum \cite{Hinshaw:2006ia};
for NVSS, the diagonal noise covariance represents shot noise and is constant between pixels.
(In both cases, a sky cut is incorporated by setting $N^{-1}$ to zero inside the mask.)
In this noise model, it is trivial to compute $N^{-1}a$ using Eq.~(\ref{eq:ninva}), 
but what we need in our analysis pipeline is $\ta = (S+N)^{-1}a$.  
Note that $N^{-1}$ is generally not invertible due to the presence of unconstrained modes
(such as pixels excluded by the sky cut), so that $a$ is not determined by Eq.~(\ref{eq:ninva}), 
but the data do determine $N^{-1}a$, and having this is sufficient for $\ta$.
The remainder of this appendix is devoted to an algorithm for computing $\ta_{\ell m}$.

Following \cite{Wandelt:2003uk}, we will find it convenient to replace the matrix $(S+N)^{-1}$ by the matrix $X^{-1}$, 
where
\ba
X &\eqdef& 1 + S^{1/2}N^{-1}S^{1/2}  \nn  \\
&=& 1 + \sum_i S^{1/2} A_i^T (N^{\rm pix}_i)^{-1} A_i S^{1/2}\,,  \label{eq:Xdef}
\ea
Using the identity $(S+N)^{-1} a = S^{-1/2}X^{-1}S^{1/2}N^{-1} a$, it suffices to give an algorithm for multiplying a
map by $X^{-1}$.
Since the number of degrees of freedom is too large for direct matrix inversion, this multiplication must be performed 
using conjugate gradient inversion \cite{NR}.
Performance of the conjugate gradient method depends on a good choice of preconditioner, or
linear operator which is efficient to compute and approximates $X^{-1}$.

A common way to construct a preconditioner is to replace $X$ by some simpler approximation $X'$
which can be inverted exactly, and use $(X')^{-1}$ as the preconditioner.
The simplest preconditioner of this type would be $X_\Delta^{-1}$,
where $X_\Delta$ is the matrix defined by keeping only the diagonal of $X$.

With this diagonal preconditioner, we have found that the conjugate gradient search will eventually converge,
but the convergence is extremely slow.
To understand why it is slow, note that $X_\Delta^{-1}$ will only be a good approximation to $X^{-1}$
when $X$ is diagonally dominated.
This will be the case on angular scales which are noise-dominated ($S/N \ll 1$), since $X$ will be close
to the identity matrix, but on large angular scales where the signal dominates, the preconditioner is not a good
approximation to $X^{-1}$, and the convergence rate becomes limited by these scales.

This picture motivates the following improved preconditioner, which has been used in several previous
treatments \cite{Hirata:2004rp,Eriksen:2004ss}.  Define the matrix $X_0$ by keeping all matrix entries in the dense block 
corresponding to multipoles $(\ell,m)$ satisfying $\ell\le\ellsplit$.  Then consider the preconditioner
\be
\left( \begin{array}{cc} X_0^{-1}  &  0  \\  0  &  X_\Delta^{-1}  \end{array}  \right)   \,,  \label{eq:blockpc}
\ee
obtained by keeping dense matrix entries below $\ellsplit$ and the diagonal above $\ellsplit$.
(In practice, the choice of $\ellsplit$ is usually dictated by memory limitations, since $\bigoh(\ellsplit^4)$
storage is needed to store $X_0$ in dense form.)
In this section, we will refer to~(\ref{eq:blockpc}) as the ``block preconditioner''.

We have found that the block preconditioner is very efficient for the NVSS dataset, but slow to converge for WMAP.
If we terminate the CG search as soon as we find an approximate solution $a' \approx X^{-1}a$
such that the termination criterion $|a - Xa'|/|a| < 10^{-6}$ is satisfied, then block preconditioning
requires $\sim 3.5$ CPU-hours to converge for the three-year WMAP dataset with Kp0 mask, and distinct beam
transfer functions for each of the eight differencing assemblies in Q, V, and W-band.

The slow convergence of this preconditioner is a bottleneck for our lens reconstruction analysis and has
also been identified as a limiting factor in other contexts, e.g. Gibbs sampling \cite{Jewell:2002dz,Wandelt:2003uk}.
Therefore, a faster method is desirable.

\subsection{Multigrid preconditioner}

\begin{figure*}
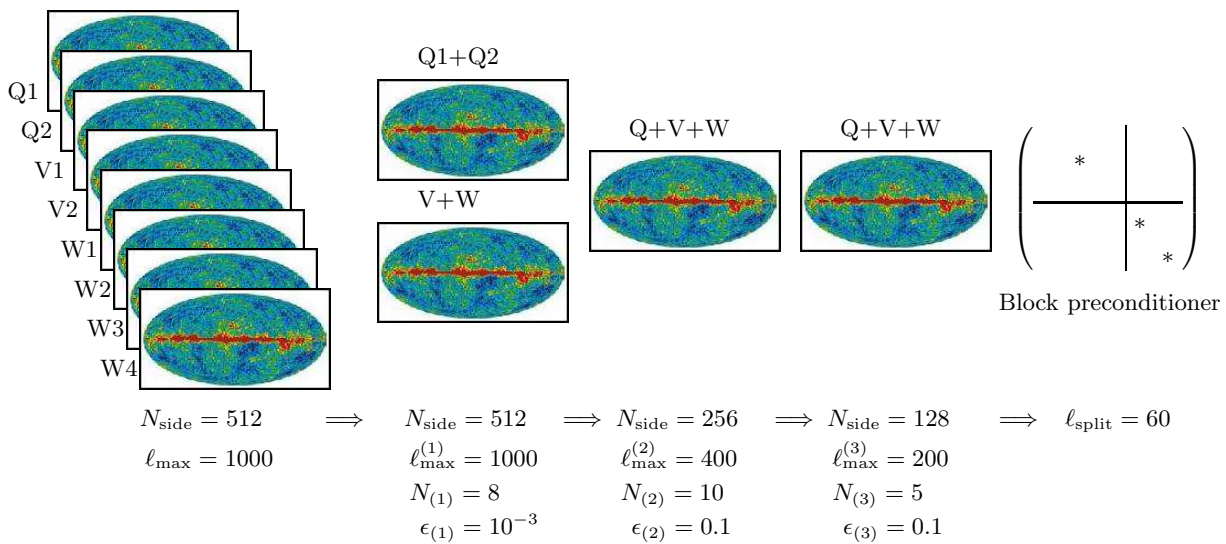

\begin{picture}(456,200)
\put(15,160){\epsfxsize=1.0truein\epsffile{wmap.eps}}
\put(20,145){\epsfxsize=1.0truein\epsffile{wmap.eps}}
\put(25,130){\epsfxsize=1.0truein\epsffile{wmap.eps}}
\put(30,115){\epsfxsize=1.0truein\epsffile{wmap.eps}}
\put(35,100){\epsfxsize=1.0truein\epsffile{wmap.eps}}
\put(40,85){\epsfxsize=1.0truein\epsffile{wmap.eps}}
\put(45,70){\epsfxsize=1.0truein\epsffile{wmap.eps}}
\put(50,55){\epsfxsize=1.0truein\epsffile{wmap.eps}}
\put(0,165){Q1}
\put(5,150){Q2}
\put(10,135){V1}
\put(15,120){V2}
\put(20,105){W1}
\put(25,90){W2}
\put(30,75){W3}
\put(35,60){W4}
\put(50,41){$\Nside=512$}
\put(52,26){$\ellmax=1000$}
\put(140,80){\epsfxsize=1.0truein\epsffile{wmap.eps}}
\put(140,134){\epsfxsize=1.0truein\epsffile{wmap.eps}}
\put(155,178){Q1+Q2}
\put(155,124){V+W}
\put(150,41){$\Nside=512$}
\put(152,26){$\ellmax^{(1)}=1000$}
\put(152,13){$N_{(1)}=8$}
\put(156,0){$\epsilon_{(1)}=10^{-3}$}
\put(220,107){\epsfxsize=1.0truein\epsffile{wmap.eps}}
\put(235,151){Q+V+W}
\put(230,41){$\Nside=256$}
\put(232,26){$\ellmax^{(2)}=400$}
\put(232,13){$N_{(2)}=10$}
\put(236,0){$\epsilon_{(2)}=0.1$}
\put(300,107){\epsfxsize=1.0truein\epsffile{wmap.eps}}
\put(315,151){Q+V+W}
\put(310,41){$\Nside=128$}
\put(312,26){$\ellmax^{(3)}=200$}
\put(312,13){$N_{(3)}=5$}
\put(316,0){$\epsilon_{(3)}=0.1$}
\put(380,125){$\left( \begin{array}{c|cc}
\makebox[0.4in]{\raisebox{0in}[0.25in][0.15in]{*}} &  &  \\  \hline
& * & \\
& & *
\end{array} \right)$}
\put(375,85){Block preconditioner}
\put(400,41){$\ellsplit=60$}
\put(120,41){$\Longrightarrow$}
\put(210,41){$\Longrightarrow$}
\put(290,41){$\Longrightarrow$}
\put(375,41){$\Longrightarrow$}
\end{picture}
\caption{Preconditioner chain for multigrid $(S+N)^{-1}$ filtering, using noise maps from the three-year WMAP dataset.
From left to right, each set of maps represents one conjugate gradient inversion problem, which is preconditioned
by the ``faster and cruder'' approximation which appears next in the chain, obtained by either reducing resolution
or the number of distinct beams retained in the problem.}
\label{fig:wmappc}
\end{figure*}

So far, we have recalled existing work in the literature: fast $(S+N)^{-1}$ filtering can be performed
via conjugate gradient inversion with the block preconditioner (Eq.~(\ref{eq:blockpc})).
In this section, we present our improvement.
The idea is that, even with the block preconditioner to do the inversion exactly at multipoles below $\ellsplit$,
conjugate gradient inversion is still limited
by the convergence rate at multipoles just above $\ellsplit$ (since the lowest multipoles will
have highest signal-to-noise).
However, these are precisely the multipoles which can be represented in a coarser pixelization.

This leads naturally to a multigrid preconditioner: one preconditions the inversion at resolution
$\Nside$ using the result of performing the inversion at coarser resolution $\Nside/2$, where
the spherical transform is faster by a factor of $\sim 8$.
This process is recursive; the inversion at resolution $\Nside/2$ is preconditioned by an
inversion at resolution $\Nside/4$, and so on.  
At the coarsest resolution (typically $\Nside=128$), the inversion is preconditioned
using the block preconditioner.
For the WMAP example with parameters as described at the end of \S\ref{ssec:cg}, we find
a running time of 14 CPU-minutes using the multigrid preconditioner.
This represents an improvement by a factor $\sim 15$, relative to the block preconditioner alone.

In detail, the multigrid method works as follows.  As described in the preceding section, we
wish to compute $X^{-1}a$, where $a = a_{\ell m}$ is defined in harmonic space up to some maximum
multipole $\ellmax$, and $X$ is defined by Eq.~(\ref{eq:Xdef}).
Then let $X_{(1)}$ be the matrix defined analagously, with all noise covariance matrices ``coarsified''
(i.e. with $\Nside$ decreased by a factor of two), 
and with the maximum multipole reduced to some $\ellmax^{(1)} < \ellmax$.
Then the multigrid preconditioner is defined by
\be
\left( \begin{array}{cc} X_{(1)}^{-1}  &  0  \\  0  &  X_\Delta^{-1}  \end{array}  \right)   \,,
\ee
i.e. we use the diagonal preconditioner for multipoles above $\ellmax^{(1)}$.
Since applying the preconditioner involves a multiplication by $X_{(1)}^{-1}$, and the matrix
$X_{(1)}$ is too large for dense inversion, we do the $X_{(1)}^{-1}$ multiplication recursively,
using an ``inner'' instance of conjugate gradient inversion.
The preconditioner for the inner CG inversion is obtained analagously by a second round of coarsifying 
noise covariance matrices and reducing the maximum multipole to some $\ellmax^{(2)} < \ellmax^{(1)}$, and so on.
At the coarsest resolution, we use the block preconditioner described in the preceding subsection.

In Figure~\ref{fig:wmappc}, we show the preconditioner chain for WMAP.
The parameters $N_{(i)}$, $\epsilon_{(i)}$ control the termination criterion for each CG instance; when
evaluating $X_{(i-1)}^{-1}$ with preconditioner $X_{(i)}^{-1}$, we terminate the CG search after $N_{(i)}$
iterations, or when the approximate solution $a' \approx X^{-1}a$ satisfies $|a - Xa'|/|a| < \epsilon_{(i)}$.
We have found that it is necessary to include these parameters to avoid spending too much CPU time in the
coarse grids.
In the WMAP3 example, the first level of preconditioning actually does not reduce the resolution,
but instead reduces the number of distinct beams in the problem from eight to two (by making the
so-called ``equal-beam approximation'' in which the average of the beam transfer functions is
used).
Note that the final output of the inversion does not make the equal-beam approximation,
but merely uses inversions with the equal-beam approximation internally, to precondition the top-level
CG inversion where no such approximation is made.

\begin{figure}
\centerline{\epsfxsize=3.2truein\epsffile[90 540 320 680]{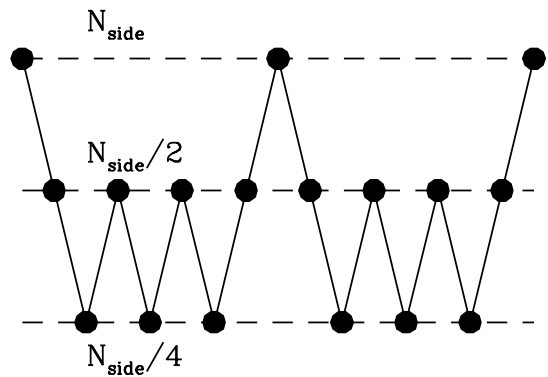}}
\caption{Sequence of coarsifying and decoarsifying operations in an instance of the multigrid
method with $N_{(1)}=3$, $N_{(2)}=2$, showing the W-cycle structure.  Each solid circle represents
one ``forward'' operation of the operator $X=(1+S^{1/2}N^{-1}S^{1/2})$ at the appropriate
resolution.}
\label{fig:wcycle}
\end{figure}

It is illuminating to describe the sequence of coarsifying and decoarsifying operations which occur
in the multigrid method.
Each iteration of the top-level CG loop requires one evaluation of its preconditioner, which in turn
is a full-blown CG search (at coarser resolution) which can iterate up to $N_{(1)}$ times.
Each of these iterations can iterate at the next coarsest resolution up to $N_{(2)}$ times, and so on.
In the parlance of multigrid algorithms, this exponential fanout is referred to as a W-cycle (Figure~\ref{fig:wcycle}).
Note that, even though the number of iterations spent at each resolution increases exponentially, the
total CPU time does not, because the running time of each iteration is exponentially supressed; in each
level, the resolution and value of $\ellmax$ are typically reduced by a factor of two, which reduces the
cost of the spherical harmonic transform by a factor of eight.
Indeed, the strength of the multigrid method is that it spends an exponentially large number of CG 
iterations on the large angular scales, which are slowest to converge but accurately approximated at
coarse resolution, while avoiding a large increase in CPU time.

The performance of the multigrid preconditioner ($\sim 14$ CPU-min per Monte Carlo WMAP simulation)
is sufficient for purposes of this paper.
However, we have also found that none of the preconditioners described so far give reasonable
performance with a realistic sky cut and the noise levels and resolution expected for the Planck
satellite mission.
Therefore, the multigrid preconditioner is probably not the final word on this subject; additional
improvements are still needed for future datasets.

\subsection{Template marginalization}
\label{ssec:complexnoise}

So far, we have assumed a noise covariance $N^{\rm map}_i$ for each map which is diagonal in pixel space.
Suppose that, in addition, one wants to marginalize the amplitudes of $\Ntmpl$ modes in the map.   We have
seen several examples in the paper:
\begin{enumerate}
\item In both WMAP and NVSS, we marginalize the monopole and dipole ($\Ntmpl=4$).
\item In NVSS, we remove systematic declination gradients by marginalizing any mode which is constant around
each isolatitude ring in equatorial coordinates (\S\ref{sec:nvss_systematics}).  
This leads to $\Ntmpl=N_{\rm ring}$, where $N_{\rm ring}$ is the number of isolatitude rings in the pixelization.
\item In WMAP, one could use this formalism to marginalize any signal proportional to external
foreground templates, although we have not implemented this because the effect of Galactic foregrounds
is small (\S\ref{ssec:galacticforegrounds}).
\end{enumerate}

Template marginalization, in this general form, is easy to incorporate in our conjugate gradient framework.
Let $\tau$ be an $\Ntmpl$-by-$\Npix$ matrix containing the templates.
By the Woodburry formula, template marginalization modifies the map covariance as follows:
\be
(N^{\rm pix}_i)^{-1} - (N^{\rm pix}_i)^{-1} \tau^T [\tau (N^{\rm pix}_i)^{-1} \tau^T]^{-1} \tau (N^{\rm pix}_i)^{-1}   \label{eq:woodburry}
\ee
Since the conjugate gradient method only requires a ``black box'' procedure for multiplying a map by the
inverse covariance $(N^{\rm pix}_i)^{-1}$, one simply includes the extra term in Eq.~(\ref{eq:woodburry}).

If $\Ntmpl$ is small (e.g. in the case of marginalizing the monopole and dipole), 
one can simply keep the matrix $\tau$ in dense form.
In cases where $\Ntmpl$ is large, all that is needed is a procedure for multiplying a map by the matrix $\tau$, 
i.e. computing each template amplitude given a map.
For example, when marginalizing declination gradients in NVSS, we implement ``multiplication by $\tau$'' by
simply averaging pixel values around each isolatitude ring in the input map.

\section{Three-point estimators}
\label{app:est}

In Appendix~\ref{app:cinv} we have described in detail how the filtered CMB map $\ta_{\ell m}$
and filtered galaxy map $\tg_{\ell m}$ are computed in our pipeline.
In order to completely describe our implementation, there is one remaining loose end: 
in this appendix, we will give the details of how our
quadratic reconstructions $\tphi_{\ell m}, \tpsi_{\ell m}, \ts_{\ell m}$ are computed.
We will also prove the statement, made throughout the paper, that our bandpower estimators
$\hC^{\phi g}_b, \hC^{\psi g}_b, \hC^{sg}_b$ for lensing, curl null test, and point sources
are optimal.
Our proof will depend on the assumption of small deviations from Gaussianity,
and we discuss the conditions under which this assumption applies.

\subsection{Quadratic reconstruction}
\label{ssec:quadrec}

Here, we give the implementational details of how the quadratic reconstuctions 
$\tphi_{\ell m}, \tpsi_{\ell m}, \ts_{\ell m}$  are computed in our pipeline.
There is a small subtlety because the reconstructions are defined by position space equations,
e.g. $\tphi_{\ell m}$ is defined by:
\be
\sum_{\ell m} \tphi_{\ell m} Y_{\ell m}(x) = \nabla^a( \alpha(x) \nabla_a \beta(x) )  \label{eq:b2}
\ee
but the maps $\ta_{\ell m}, \tphi_{\ell m}$ are defined in harmonic space.
(The quantities $\alpha(x), \beta(x)$ were defined in Eqs.~(\ref{eq:alphadef}),~(\ref{eq:betadef}).)

In principle, $\tphi$ can be evaluated as a brute force harmonic space sum:
\be
\tphi_{\ell m}^* = \sum_{\ell_1m_1\ell_2m_2} f_{\ell_1\ell\ell_2} C^{TT}_{\ell_2}
                                \gau^{\ell\ell_1\ell_2}_{mm_1m_2} \ta_{\ell_1m_1} \ta_{\ell_2m_2}  \label{eq:bfphi}
\ee
where $f_{\ell_1\ell_2\ell_3}$ was defined previously in Eq.~(\ref{eq:fdef}), and we have introduced the notation
\ba
\gau^{\ell_1\ell_2\ell_3}_{m_1m_2m_3} &\eqdef& \sqrt{\frac{(2\ell_1+1)(2\ell_2+1)(2\ell_3+1)}{4\pi}}  \nn  \\
   && \times \threej{\ell_1}{\ell_2}{\ell_3}{0}{0}{0}  \threej{\ell_1}{\ell_2}{\ell_3}{m_1}{m_2}{m_3} 
\ea
However, the harmonic-space sum has computational cost $\bigoh(\ellmax^5)$ and so we introduce
an optimized position-space method.

Multiplying Eq.~(\ref{eq:b2}) on both sides by $Y_{\ell m}(x)^*$ and integrating over $x$, one obtains:
\be
\tphi_{\ell m} = \int d^2x\, \nabla^a(Y_{\ell m}(x))^* \alpha(x) \nabla_a \beta(x)  \label{eq:b1}
\ee
The integral can be done exactly using Gauss-Legendre quadrature in $\cos(\theta)$ and
uniform quadrature in $\varphi$.
We evaluate the quantities $\alpha(x), \nabla_a\beta(x)$ on the isolatitude rings by
using a fast spin-0 and spin-1 spherical transform respectively.
The right-hand side of Eq.~(\ref{eq:b1}) can then be evaluated using a fast spin-1 transform.
This algorithm provides an exact evaluation of Eq.~(\ref{eq:b1}) with computational cost $\bigoh(\ellmax^3)$.
We use an analagous method to evaluate the quadratic quantities $\tpsi_{\ell m}, \ts_{\ell m}$.

\subsection{Equivalence with the bispectrum}

As a preliminary step toward proving optimality,
we show how the estimators $\hC_b^{\phi g}, \hC_b^{\psi g}, \hC_b^{sg}$
can be rewritten purely in terms of the associated bispectra.
Throughout this paper, when we write a bispectrum $b_{\ell_1\ell_2\ell_3}$, it is understood
that $\ell_1,\ell_2$ are CMB multipoles and $\ell_3$ is a galaxy multipole.

We write the lensing estimator in the following form:
\be
\hC^{\phi g} = \frac{1}{\mathcal N} \sum_{\ell m}  C_\ell^{\phi g}
                      [ \tphi_{\ell m} - \langle\tphi_{\ell m}\rangle ]^* 
                     \tg_{\ell m}    \label{eq:bispec1}
\ee
In Eq.~(\ref{eq:bispec1}) and throughout this appendix, $C_\ell^{\phi g}$ denotes the cross power spectrum
we are interested in estimating (typically proportional to $1/\ell^2$ over some band in $\ell$), not the
fiducial spectrum.

If we replace $\tphi_{\ell m}^*$ by the right-hand side of Eq.~(\ref{eq:bfphi}), we obtain:
\ba
\hC_b^{\phi g} &=& \frac{1}{\mathcal N} \sum_{\ell_im_i}
 f_{\ell_1\ell_2\ell_3} C_{\ell_2}^{TT} C_{\ell_3}^{\phi g} \gau^{\ell_1\ell_2\ell_3}_{m_1m_2m_3}   \\
&& \qquad\times \big[\ta_{\ell_1m_1}\ta_{\ell_2m_2}-\langle\ta_{\ell_1m_1}\ta_{\ell_2m_2}\rangle \big] \tg_{\ell_3m_3} \nn
\ea
We can replace
$\langle \ta_{\ell_1m_1} \ta_{\ell_2m_2}\rangle$ by $C^{T\,-1}_{\ell_1m_1,\ell_2m_2}$,
where in this appendix
we use the notation $(C^T)^{-1}, (C^g)^{-1}$ to distinguish the
inverse signal + noise covariances for the CMB and galaxy fields.
Now comparing
with the form of the bispectrum due to lensing (Eq.~(\ref{eq:Blensing})), this becomes:
\ba
\hC^{\phi g}_b &=& \frac{1}{2 \mathcal N}
\sum_{\ell_i m_i} b_{\ell_1\ell_2\ell_3} \gau^{\ell_1\ell_2\ell_3}_{m_1m_2m_3}  \label{eq:bispec2}  \\
&& \qquad\qquad\times \big[ \ta_{\ell_1m_1} \ta_{\ell_2m_2} - C^{T\,-1}_{\ell_1m_1,\ell_2m_2} \big] \tg_{\ell_3m_3}  \nn
\ea
We have now written the lensing estimator purely in terms of the lensing bispectrum $b_{\ell_1\ell_2\ell_3}$.
A similar calculation shows that the same is true for the curl and point source estimators 
$\hC_b^{\psi g}, \hC_b^{sg}$: in both cases the estimator takes the form in Eq.~(\ref{eq:bispec2}), 
with $b_{\ell_1\ell_2\ell_3}$
replaced by the bispectrum due to lensing by a curl component, or the point source bispectrum in Eq.~(\ref{eq:bell3}).
This allows us to give a uniform proof of optimality which applies to all three cases, as we will now see.

\subsection{Optimality}

We will now prove the following general statement: for {\em any} bispectrum
$b_{\ell_1\ell_2\ell_3}$, the optimal estimator is given by
\be
\lest = \frac{1}{F} (\lest_3 - \lest_1)   \label{eq:optest}
\ee
where the three-point and one-point terms are defined by
\ba
\lest_3 &\eqdef& \frac{1}{2} \sum_{\ell_im_i} b_{\ell_1\ell_2\ell_3} \gau^{\ell_1\ell_2\ell_3}_{m_1m_2m_3}
               \ta_{\ell_1 m_1} \ta_{\ell_2 m_2} \tg_{\ell_3 m_3}  \\
\lest_1 &\eqdef& \frac{1}{2} \sum_{\ell_im_i} b_{\ell_1\ell_2\ell_3} \gau^{\ell_1\ell_2\ell_3}_{m_1m_2m_3}
               C^{T\, -1}_{\ell_1m_1,\ell_2m_2} \tg_{\ell_3m_3}
\ea
and $F$ is the Fisher matrix element
\ba
F &\eqdef& \frac{1}{2} \sum_{\ell_i m_i}
b_{\ell_1\ell_2\ell_3}
b_{\ell_4\ell_5\ell_6}
\gau^{\ell_1\ell_2\ell_3}_{m_1m_2m_3}
\gau^{\ell_4\ell_5\ell_6}_{m_4m_5m_6}  \times  \nn \\
&& \qquad
C^{T\,-1}_{\ell_1m_1,\ell_4m_4}
C^{T\,-1}_{\ell_2m_2,\ell_5m_5}
C^{g\,-1}_{\ell_3m_3,\ell_6m_6}  \label{eq:fishergeneral}
\ea
(This expression generalizes the Fisher matrix for all sky isotropic noise previously considered
in Eq.~(\ref{eq:fisheriso}) to an arbitrary noise covariance.)
Note that we have computed the normalization explicitly; a short calculation shows that the
estimator in Eq.~(\ref{eq:optest}) has unit response to the bispectrum $b_{\ell_1\ell_2\ell_3}$, so that
the estimator is normalized and does not need the $1/{\mathcal N}$ prefactor.

The proof will depend on the assumption of weak non-Gaussianity; specifically we will assume that the
fields are sufficiently close to Gaussian that the estimator variance can be approximated by its Gaussian
contribution.

First, we can show using the Cramer-Rao inequality that any unbiased estimator ${\mathcal E}$ has
variance $\ge 1/F$, where $F$ is the Fisher matrix in Eq.~(\ref{eq:fishergeneral}).
This is proved using the method of \cite{Babich:2005en,Creminelli:2005hu}, expanding the likelihood
function for $a_{\ell m}, g_{\ell m}$ around its Gaussian limit using the Edgeworth expansion.

Now consider the variance $\Var(\hC)$.  We are assuming that this variance may be calculated
using Gaussian statistics, so that Wick's theorem gives:
\ba
\Var(\lest_3,\lest_3) &=& F + f^T (C^g)^{-1} f  \\
\Cov(\lest_3,\lest_1) &=& \Cov(\lest_1,\lest_1) = f^T (C^g)^{-1} f  \nn
\ea
where we have defined
\ba
f_{\ell m} = \frac{1}{2} \sum_{\ell_im_i} b_{\ell_1\ell_2\ell} 
   \gau^{\ell_1\ell_2\ell}_{m_1m_2m} C^{T\,-1}_{\ell_1m_1,\ell_2m_2}
\ea
Putting this together, we get $\Var(\lest) = 1/F$, i.e. the Cramer-Rao inequality is saturated.
This completes our proof that the estimator is optimal, under the assumption of weak non-Gaussianity.

When is this assumption satisfied for lensing?
Roughly speaking, weak non-Gaussianity starts to break down when the
instrumental sensitivity becomes good enough that a high signal-to-noise detection of
CMB lensing can be achieved.
More precisely, consider the case of full sky coverage and isotropic noise.  This allows
us to make contact with the results of \cite{Okamoto:2003zw}, where an unbiased estimator $\hphi_{\ell m}$
is defined for each multipole of the lensing potential, with full-sky noise power spectrum
$N_\ell^{\phi\phi}$ given previously in Eq.~(\ref{eq:nlpp}).
In this notation, one can show that our filtered field $\tphi_{\ell m}$ is equal to 
$(N_\ell^{\phi\phi})^{-1} \hphi_{\ell m}$, and our estimator is given by:
\be
\lest = \sum_{\ell m} C_\ell^{\phi g} 
                          \left( \frac{\hphi_{\ell m}^*}{N_\ell^{\phi\phi}} \right)
                          \left( \frac{g_{\ell m}}{C_\ell^{gg} + N_\ell^{gg}} \right)
\label{eq:lensing_estimator}
\ee
The first improvement that can be made to this estimator is to make the replacement
\be
\lest \rightarrow \lest' =  \sum_{\ell m} C_\ell^{\phi g} 
                          \left( \frac{\hphi_{\ell m}^*}{C_\ell^{\phi\phi} + N_\ell^{\phi\phi}} \right)
			  \left( \frac{g_{\ell m}}{C_\ell^{gg} + N_\ell^{gg}} \right)
\ee
to incorporate the nonzero sample variance of the lenses. 
Our estimator $\lest$ is optimized assuming Gaussian covariance among modes
of the CMB, and does not ``know'' that there is extra sample variance hidden in the problem.
However, it is unclear how to generalize $\lest'$ to the case of sky cuts and inhomogeneous
noise, as we have done for $\lest$, allowing an arbitrary noise covariance matrix $N$.

The estimators $\lest,\lest'$ agree when $C_\ell^{\phi\phi} \ll N_\ell^{\phi\phi}$, i.e. when
the reconstruction noise in the lensing potential dominates the signal, considered one
mode of the potential at a time.
This condition holds for WMAP, as can be seen from the direct comparison in 
Fig.~\ref{fig:reconstruction_noise}, left panel.
However, the estimator $\lest$ which we have constructed would start to become suboptimal
for future surveys with sufficient sensitivity to reconstruct the lensing potential with
signal-to-noise $\sim 1$ per mode.
For even more futuristic sensitivity levels, even the improved estimator $\lest'$ would
become suboptimal; the three-point estimator could be improved by using a maximum likelihood
formalism which incorporates information from higher-point correlation functions 
of all orders \cite{Hirata:2002jy}.

In addition to these optimality issues for future surveys,
there are other ways in which our estimator might be extended.
First, we have not considered CMB polarization, which is ultimately expected to provide more
sensitivity to lensing than temperature \cite{Hu:2001kj}.
Second, by using full-blown $C^{-1}$ filtering, we have ensured optimality of the estimator,
but it would be interesting to determine whether a simpler filter can be found which achieves
near-optimal power spectrum uncertainties.
As we have remarked in Appendix~\ref{app:cinv}, the $C^{-1}$ operation seems prohibitively expensive
for Planck with existing preconditioners, so finding such a filter may be a practical necessity
for future experiments.

\section{Resolved point sources}
\label{app:mask}

We give a proof of a statement made in \S\ref{ssec:mask}:
correlations between the mask and the galaxy field cannot fake
the lensing signal, i.e. the expectation value
\be
\left\langle \hC_b^{\phi g} \right\rangle_{T,G,M} = 0  \label{eq:fakesignal}
\ee
in the absence of CMB lensing.
We have introduced the notation $\langle \cdot \rangle_{T,G,M}$ to denote an expectation
value taken over realizations of the CMB $T$, galaxy counts $G$, and mask $M$
(where the last two are assumed correlated).

In the proof, we will denote the quadratic reconstruction $\tphi$ which appears in the
lensing estimator by $\tphi(T,M)$ to emphasize that it depends on both
the CMB realization $T$ and the mask $M$.  We will analagously denote the filtered
galaxy field by $\tg(G,M)$.  In this notation, the lensing estimator can be written:
\be
\hC_b^{\phi g} = \sum_{\ell m} \Big[ \tphi(T,M) - \langle \tphi(T',M) \rangle_{T'} \Big]^*_{\ell m} \tg(G,M)_{\ell m}
\ee
where we have written the one-point term as an average over CMB realization $T'$ with the mask $M$ fixed.
Taking the expectation value $\langle\cdot\rangle_{T,G,M}$ on both sides, we obtain:
\ba
\left\langle \hC_b^{\phi g} \right\rangle_{T,G,M} &=&
\Bigg\langle \sum_{\ell m} \big\langle \tphi(T,M)^*_{\ell m} \big\rangle_T \big\langle \tg(G,M)_{\ell m} \big\rangle_G \nn  \\
&& -  \big\langle \tphi(T',M)^*_{\ell m} \big\rangle_{T'} \big\langle \tg(G,M)_{\ell m} \big\rangle_G \Bigg\rangle_M \nn \\
&=& 0
\ea
In the first line, we have used the fact that in the absence of lensing, the CMB realization $T$ is independent of
the galaxy realization $G$ once the mask $M$ has been specified, to bring the expectation value $\langle\cdot\rangle_T$
inside the sum.
This completes the proof that the expectation value in Eq.~(\ref{eq:fakesignal}) vanishes in the
absence of CMB lensing, i.e. mask-galaxy correlations cannot fake the lensing signal.

It is interesting to note that this proof would break down if the one-point term were omitted
from the lensing estimator $\hC_b^{\phi g}$.  In this case, we would obtain
\be
\langle \hC_b^{\phi g} \rangle_{T,g,M} = 
\left\langle \sum_{\ell m} \langle \tphi(T,M)_{\ell m}^* \rangle_T \langle \tg(G,M)_{\ell m} \rangle_G \right\rangle_M
\ee
which cannot be simplified further: the map $\langle\tg(G,M)\rangle_G$ can be nonzero if there are mask-galaxy
correlations, and the map $\langle\tphi(T,M)\rangle_T$ is generally nonzero in the presence of a mask.

\section{Beam asymmetry}
\label{app:asymm}

To include beam asymmetry in our simulation pipeline, we need an expression
for the beam-convolved temperature $\tT(x)$ in each pixel $x$, in terms of three
quantities: the beam profile, the scan strategy, and the unconvolved CMB $T(x)$.

We represent the beam profile in real space as $G(\theta,\varphi)$ or in harmonic
space as:
\be
G(\theta,\varphi) = \sum_{\ell s} g_{\ell s} Y_{\ell s}(\theta,\varphi)  \label{eq:GFT}
\ee
Following \cite[Appendix B]{Hinshaw:2006ia}, the scan strategy will be represented by the following quantity:
\be
w(x,\alpha) = 2\pi \frac{\sum_{i\in x} \delta(\alpha-\alpha_i)}{\sum_{i\in x} 1}  \label{eq:wdef}
\ee
where the angle $\alpha$ parameterizes beam orientations at the pixel $x$, relative to an
arbitrarily chosen reference angle.
The sum in Eq.~(\ref{eq:wdef}) runs over timestream samples $i$ which fall in pixel $x$ with beam 
orientation $\alpha_i$.
Note that $w(x,\alpha)$ depends on the choice of reference direction, or local frame at $x$.

We briefly recall the theory of spin-$s$ fields; for more details see \cite{Zaldarriaga:1996xe}.
A spin-$s$ field ($-\infty < s < \infty$) is a function $({}_sf)$
whose value at $x$ depends on a choice of frame, or pair of orthonormal basis vectors
$\{ \e_1, \e_2 \}$ at $x$.  Under the right-handed rotation
\begin{eqnarray}
\e'_1 &=&  (\cos\theta)\e_1 + (\sin\theta)\e_2    \\
\e'_2 &=& -(\sin\theta)\e_1 + (\cos\theta)\e_2   \nonumber
\end{eqnarray}
$({}_sf)$ must transform as $({}_sf)' = e^{-is\theta} ({}_sf)$.
One can define spin-$s$ spherical harmonics ${}_sY_{\ell m}(\theta,\varphi)$; these are an orthonormal set of
basis functions for spin-$s$ fields, with properties that are similar to the ordinary (spin-0) spherical
harmonics $Y_{\ell m}$.

If we Fourier transform the frame-dependent quantity $w(x,\alpha)$ in the angle $\alpha$:
\be
w(x,\alpha) = \sum_{s=-\infty}^{\infty} ({}_sw(x))^* e^{is\alpha}  \label{eq:wFT}
\ee
then ${}_sw(x)$ will be a spin-$s$ field as suggested by the notation.

Now we can write an expression for the beam-convolved CMB temperature $\tT(x)$:
\be
\tT(x) = \int d^2x'\, T(x') \int \frac{d\alpha}{2\pi}\, w(x,\alpha)_{\rm 2P}\, G(\theta_{xx'},-\alpha)  \label{eq:beam1}
\ee
where $\theta_{xx'}$ denotes the angle between points $x,x'$, and
the subscript ``2P'' on any frame-dependent quantity (such as $w(x,\alpha)$) indicates the
``two-point'' frame: the reference direction $\e_1$ at $x$ points toward $x'$.

Eq.~(\ref{eq:beam1}) simply states that the beam-convolved temperature at $x$ is given by averaging over scan
directions $\alpha$, with the beam profile rotated through angle $\alpha$ before it is applied.
To simplify this expression, we plug in Eqs.~(\ref{eq:GFT}),~(\ref{eq:wFT}), obtaining:
\be
\tT(x) = \int d^2x\, T(x) \sum_{s\ell} ({}_sw(x)_{\rm 2P})^* g_{\ell s} Y_{\ell s}(\theta_{xx'},0)
\ee
Now use the identity
\be
Y_{\ell s}(\theta_{xx'},0) = \sqrt{\frac{4\pi}{2\ell+1}} \sum_m ({}_{s}Y_{\ell m}(x))_{\rm 2P}\, Y^*_{\ell m}(x')
\ee
to obtain
\be
\tT(x) = \sum_{s\ell m} \sqrt{\frac{4\pi}{2\ell+1}} ({}_sw(x))^* g_{\ell s} a_{\ell m} ({}_sY_{\ell m}(x)) \label{eq:beamf}
\ee
This is our desired expression for $\tT$.  
The final result is a spin-0 quantity, so we have dropped the 2P.

Eq.~(\ref{eq:beamf}) is a sum over beam multipoles $s$ multiplied by the spin-$s$ component of the scan strategy.
Note that the spin-0 component $({}_0w(x))$ is equal to 1 by construction (Eq.~(\ref{eq:wdef})), so that the $s=0$
term in Eq.~(\ref{eq:beamf}) does not depend on the scan strategy and is simply given by convolving $\{a_{\ell m}\}$
with the beam transfer function $b_\ell = \sqrt{4\pi/(2\ell+1)} g_{\ell 0}$.
The higher-spin terms do depend on the scan strategy and represent corrections to the symmetric-beam
approximation.
If the beam is azimuthally symmetric ($g_{\ell s}=0$ for $s>0$), or the beam is arbitrary but the scan
is isotropic in each pixel (${}_sw(x)=0$ for $s>0$), then the higher spin terms do not contribute and the
symmetric-beam approximation is exact.
For WMAP, we find that the sum over $s$ in Eq.~(\ref{eq:beamf}) converges rapidly so that truncating at $\smax=16$
fully incorporates beam asymmetry.

A fast algorithm for evaluating Eq.~(\ref{eq:beamf}) may be given by noting that each term in the $s$ sum is
simply a spin-$s$ spherical transform.  In an isolatitude coordinate system, a spin-$s$ transform may
be performed with computational cost $\bigoh(\ellmax^3)$ by using the recursion
\be
\rho^s_{\ell m} ({}_sY_{\ell m}) = 
\left(z+\frac{sm}{\ell(\ell-1)}\right) {}_sY_{\ell-1,m} -
\rho^s_{\ell-1,m} ({}_sY_{\ell-2,m}) 
\ee
on each isolatitude ring,
where we have defined $\rho^s_{\ell m} = \sqrt{(\ell^2-m^2)(\ell^2-s^2)/(4\ell^2-1)}/\ell$.
Thus the total computational cost of incorporating beam asymmetry via Eq.~(\ref{eq:beamf}) is $\bigoh(\smax\ellmax^3)$.

Finally, we include a detail which is specific to WMAP.  The preceding treatment has assumed that there is one
beam $g_{\ell s}$ and one scan ${}_sw(x)$ for each simulated map.  In WMAP, we have one simulated map per differencing
assembly, obtained as the difference of A-side and B-side measurements.  In this case, one makes the replacement
\be
({}_sw(x))^* g_{\ell s} \rightarrow ({}_sw^A(x))^* g^A_{\ell s} + ({}_sw^B(x))^* g^B_{\ell s}
\ee
in Eq.~(\ref{eq:beamf}),
where $g^A_{\ell m}, g^B_{\ell m}$ are the A-side and B-side beams, and $w^A,w^B$ are defined by
\ba
w^A(x,\alpha) & \eqdef & 2\pi 
\frac{\sum_{a\in x} \delta(\alpha-\alpha_a)}{\left(\sum_{a\in x} 1\right) + \left( \sum_{b\in x} 1 \right)}  \\
w^B(x,\alpha) & \eqdef & 2\pi 
\frac{\sum_{b\in x} \delta(\alpha-\alpha_b)}{\left(\sum_{a\in x} 1\right) + \left( \sum_{b\in x} 1 \right)}
\ea
where $\sum_{a\in x},\sum_{b\in x}$ denote sums over A-side and B-side timestream samples which fall in
pixel $x$.

\vfill
\bibliography{wmap-nvss}

\end{document}